\documentclass[useAMS,usenatbib,usegraphicx]{mn2e}

\newcommand{\gapprox}{\,\rlap{\lower 2.5pt 
\hbox{$\sim$}}\raise 1.5pt\hbox{$>$}\,}
\newcommand{\gsim}{\,\rlap{\lower 2.5pt 
\hbox{$\sim$}}\raise 1.5pt\hbox{$>$}\,}
\newcommand{\lapprox}{\,\rlap{\lower 2.5pt 
\hbox{$\sim$}}\raise 1.5pt\hbox{$<$}\,}
\newcommand{\lsim}{\,\rlap{\lower 2.5pt 
\hbox{$\sim$}}\raise 1.5pt\hbox{$<$}\,}

\newcommand{\chiquadr}{\ensuremath{\chi^{2}_{\rm r}}}

\def\eeq{\end{equation}}
\def\beq{\begin{equation}}

 \usepackage{times}

\title[BOSS stellar masses]{Stellar masses of SDSS-III/BOSS galaxies at $z\sim0.5$~and constraints to galaxy formation models}
\author[C. Maraston et al.]{Claudia Maraston$^{1,4}$\thanks{E-mail:
claudia.maraston@port.ac.uk}, Janine Pforr$^{2}$, Bruno M. Henriques$^{3}$, Daniel Thomas$^{1,4}$, 
\newauthor David Wake$^{5}$, Joel R. Brownstein$^{14}$, Diego Capozzi$^{1,4}$, Jeremy Tinker$^{6}$, Kevin Bundy$^{7}$, 
\newauthor Ramin A. Skibba$^{8}$, Alessandra Beifiori$^{9}$, Robert C. Nichol$^{1,4}$, Edd Edmondson$^{1,4}$, 
\newauthor Donald P. Schneider$^{10,11}$, Yanmei Chen$^{12,13}$, Karen L. Masters$^{1,4}$, Oliver Steele$^{1,4}$, 
\newauthor Adam S. Bolton$^{14}$, Donald G. York$^{15}$, Benjamin A. Weaver$^{16}$, Tim Higgs $^{1,4}$,Dmitry Bizyaev$^{17}$, 
\newauthor Howard Brewington$^{17}$, Elena  Malanushenko$^{17}$, Viktor Malanushenko$^{17}$, Stephanie Snedden$^{17}$, 
\newauthor Daniel Oravetz$^{17}$, Kaike Pan$^{17}$, Alaina Shelden$^{17}$, Audrey Simmons$^{17}$\\
$^{1}$Institute of Cosmology and Gravitation, Dennis Sciama Building, Burnaby 
Road, Portsmouth PO1 3FX\\
$^{2}$National Optical Astronomy Observatories, 950 N. Cherry Ave., Tucson, AZ, 85719, USA\\
$^{3}$Max-Planck-Institut f\"ur Astrophysik, Karl-Schwarzschild-Str. 1,
85741 Garching b. M\"unchen, Germany\\
$^{4}$South East Physics Network, www.sepnet.ac.uk\\
$^{5}$Department of Astronomy,Yale University,P.O. Box 208101, New Haven, CT 06520-8101, USA\\
$^{6}$Center for Cosmology and Particle Physics,Department of Physics, New York University, NYC, USA\\
$^{7}$Kavli Institute for the Physics and Mathematics of the Universe, Todai Institutes for Advanced
Study, the University of Tokyo, Kashiwa, Japan 277-8583 (Kavli IPMU, WPI)\\
$^{8}$Steward Observatory, University of Arizona, 933 North Cherry Avenue,
Tucson, AZ 85721, USA\\
$^{9}$Max-Planck-Institut f\"ur Extraterrestrische Physik, Giessenbachstra§e, D-85748 Garching, Germany\\
$^{10}$Department of Astronomy and Astrophysics, The Pennsylvania State University, University Park, PA 16802\\
$^{11}$Institute for Gravitation and the Cosmos, The Pennsylvania State University, University Park, PA 16802\\
$^{12}$ Department of Astronomy, Nanjing University, Nanjing 210093, China\\ 
$^{13}$ Key Laboratory of Modern Astronomy and  Astrophysics (Nanjing University), Ministry of Education, Nanjing 210093, China\\
$^{14}$Department of Physics \& Astronomy, University of Utah, 115 South 1400 East, Salt Lake City, UT 84112 USA\\
$^{15}$Department of Astronomy and Astrophysics and the Fermi Institute, The University of Chicago, 5640 South Ellis Avenue, Chicago, IL 60615\\
$^{16}$Center for Cosmology and Particle Physics, New York University, New York, NY 10003 USA
$^{17}$Apache Point Observatory, P.O. Box 59, Sunspot, NM 88349-0059, USA
}
\begin{document}
\bibliographystyle{mn2e}
\date{Accepted. Received; in original form 25 July 2012}
\pagerange{\pageref{firstpage}--\pageref{lastpage}} \pubyear{2012}
\maketitle
\label{firstpage}
\begin{abstract}
We calculate stellar masses for $\sim 400,000$ massive luminous galaxies at redshift $\sim 0.2-0.7$~using the first two years of data from the Baryon Oscillation Spectroscopic Survey (BOSS). Stellar masses are obtained by fitting model spectral energy distributions to $u,g,r,i,z$~magnitudes, and simulations with mock galaxies are used to understand how well the templates recover the stellar mass.
Accurate BOSS spectroscopic redshifts are used to constrain the fits. We find that the distribution of stellar masses in BOSS is narrow ($\Delta~\log M\sim0.5$~dex) and peaks at about $\log~M/M_{\odot}\sim11.3$ (for a Kroupa initial stellar mass function), and that the mass sampling is uniform over the redshift range 0.2 to 0.6, in agreement with the intended BOSS target selection. 
The galaxy masses probed by BOSS extend over $\sim 10^{12} M_{\odot}$, providing unprecedented measurements of the high-mass end of the galaxy mass function. We find that the galaxy number density above $\sim 2.5\cdot 10^{11} M_{\odot}$ agrees with previous determinations. 
We perform a comparison with semi-analytic galaxy formation models tailored to the BOSS target selection and volume, in order to contain incompleteness. The abundance of massive galaxies in the models compare fairly well with the BOSS data, but the models lack galaxies at the massive end. Moreover, no evolution with redshift is detected from $\sim 0.6$ to 0.4 in the data, whereas the abundance of massive galaxies in the models increases to redshift zero. Additionally, BOSS data display colour-magnitude (mass) relations similar to those found in the local Universe, where the most massive galaxies are the reddest. On the other hand, the model colours do not display a dependence on stellar mass, span a narrower range and are typically bluer than the observations. We argue that the lack of a colour-mass relation for massive galaxies in the models is mostly due to metallicity, which is too low in the models. 
\end{abstract}
\section{Introduction}
In the cold dark matter hierarchical Universe model \citep{whiree78}, galaxies grow from primordial density fluctuations in the power spectrum \citep{bluetal84,davetal85} and assemble their mass over cosmic time through a variety of processes, such as star formation, merging and accretion \citep[e.g.][]{kauetal93,sompri99,coletal00,hatetal03,menetal04,monfontaf07,hentho10,guoetal11,henetal12}. The observational tracing of the galaxy mass growth as a function of redshift is a powerful diagnostic of the galaxy formation process, which has been investigated by many groups, through large galaxy surveys (e.g. the Sloan Digital Sky Survey, SDSS, York et al. 2000; COMBO-17, Wolf et al. 2001; MUNICS, Drory et al. 2001; DEEP2, Davis et al. 2003; GOODS, Dickinson et al. 2003; VVDS, Le F\'evre et al. 2005; 2SLAQ, Cannon et al. 2006; COSMOS, Scoville et al. 2007; GMASS, Kurk et al. 2008; GAMA, Driver et al. 2011; CANDELS, Grogin et al. 2011, Koekemoer et al. 2011; SERVS, Mauduit et al. 2012. See also the review by Renzini 2006)\nocite{yoretal00,woletal01,droetal01,davetal03,dicetal03,lefetal05,canetal06,scoetal07,kuretal08,drietal11,groetal11,mauetal12,ren06}. 

The massive (M$\gapprox 5\cdot10^{10}~\rm M_{\odot}$) component of the galaxy population is particularly interesting in the context of galaxy formation and cosmology because the stellar population properties, such as stellar ages and chemical abundances, of massive galaxies are notoriously challenging to models, e.g. the high-fraction of $\alpha$-elements over iron and the [$\alpha$/Fe] versus galaxy stellar mass relation \citep{woretal92,davsadpel93,cardan94,rosetal94,benpaq95,joretal95,gre97,traetal00,kun00,prosan02,smietal09,thoetal05,thoetal10}, the total stellar metallicity and its dependence on stellar mass, which we shall focus on in this paper \citep{pipetal09,hentho10,saketal11,delbor12}; the uniformly old stellar ages with little evidence of star formation \citep{bowlucel92,bowkodter98,thoetal05,beretal06}, the independence of the stellar population properties of the environment \citep{penetal10,thoetal10}. There are still many unknowns in the process of galaxy formation and evolution, both at the high and low mass end of the galaxy distribution (see reviews by Silk 2011 and White 2011)\nocite{sil11,whi11},  which are thought to be mostly related to the baryonic component of galaxies, especially to the poorly known processes involving gas physics, such as star formation and feedback from stars and AGN \citep[e.g.][]{govetal98,kauhae00,croetal06,bowetal06,cioost07,oppdav08,johetal12}, and their interplay with the mass assembly over cosmic time \citep[e.g.,][]{bowetal12}.

An efficient way to probe the galaxy formation process is to study the galaxy luminosity and stellar mass functions and their evolution with redshift. In the local universe, recent results on the stellar mass function of galaxies include \citet{blaetal03}, \citet{beletal03}, \citet{baletal04}, \citet{baletal06}, \citet{baletal08}, \citet{liwhi09}, \citet{baletal12}.

At larger look-back times, several authors studied the stellar mass function as a function of redshift \citep{briell00,droetal01,droetal04,droetal05,coh02,dicetal03,fonetal03,fonetal06,rudetal03,glaetal04,bunellcon05,conetal05,boretal06,cimdadren06,bunetal06,pozetal07,pergonetal08,maretal09,ilbetal10,pozetal10}, reaching redshifts of about 4. At $z<1$, which is the focus of this work, the galaxy stellar mass function appears to evolve slowly, with about half of the total stellar mass density at $z\sim 0$ already in place at $z\sim 1$. Moreover, little if no evolution is detected at the high-mass end ($M\gapprox 10^{11}~M_{\odot}$), which is one of the manifestations of the {\it downsizing} scenario for galaxy formation in both star formation and mass assembly \citep{cimdadren06,ren06,ren09,penetal10}. Such limited evolution for the most massive galaxies below $z\sim1$ is also supported by luminosity function studies \citep{waketal06,cooetal08} as well as by the lack of evolution of galaxy clustering \citep{waketal08,tojper10}.

In this work we exploit the Baryon Oscillation Spectroscopic survey (BOSS; Schlegel et al. 2009; Dawson et al. 2013)\nocite{schwhieis10}, which is part of the Sloan Digital Sky Survey III \citep{eisetal11}, for calculating galaxy stellar masses and the galaxy stellar mass function at $z\sim0.5$. The advantage offered by BOSS is the unprecedented survey area - 10,000 square $\deg$ in total, and roughly 1/3 complete at the time of writing - and a selection cut favouring the most massive galaxies ($M\gapprox 10^{11}~M_{\odot}$). The huge area coverage, and the redshift range, which lies in the middle of the theoretical late-time mass-assembly epoch \citep{deletal06}, renders BOSS an excellent survey for galaxy evolution studies.

In this first study we do not apply completeness corrections and focus on a {\it light-coned} mass function. The comparison with galaxy formation models will be performed with simulations tailored to the BOSS target selection and volume. The global stellar mass and luminosity function for the BOSS survey, including completeness, will be published in subsequent papers. As we will see from the comparison with other published mass functions, BOSS may be essentially complete at the high-mass end (M$\gapprox 5\cdot10^{11}~M_{\odot}$).

The aim of this publication is twofold. First, we describe the stellar mass calculation and discuss the results. We also compare photometric masses with spectroscopic ones that were obtained using PCA algorithm applied to BOSS spectra \citep{cheetal12}. We then calculate the mass function over the redshift range 0.45 to 0.7 and compare the resulting stellar mass density and the galaxy colours with semi-analytic models of galaxy formation and evolution, to obtain clues to the late-time evolution of massive galaxies. In particular, given the unprecedented statistics offered by the BOSS sample at the massive end, we can study whether the main body of passive galaxies in the models has the correct mass distribution and the right colours. 

There have been several examples of such an approach in the literature. \citet{benetal03} extensively studied the constraints to the theoretical galaxy luminosity function that are posed by data in the local Universe. \citet{almetal08} focus on luminous, red galaxies at $z\lapprox~0.5$~ and compare the observed luminosity function with galaxy formation models - by Bower et al. (2006) and \citet{bauetal05} - which adopt different feedback mechanisms for quenching star formation. \citet{fonetal09} study the comparison of the stellar mass function in various semi-analytic models with data over a wide redshift range. \citet{neiwei10} discuss degeneracies of semi-analytic models including different prescriptions for cooling and feedback, and their ability to match several observational constraints, including the galaxy mass function. The task of comparing galaxy formation models to quantities derived from data, especially at high look-back time, is not an easy one, as modelled data rather than pure observables need to be used. Some works have concentrated on the {\it observed-frame} which avoids the extra-assumptions involved in translating the observed colours and luminosities into physical quantities \citep{tonetal09}, while others support the use of the {\it derived-property} plane in any case \citep{conwhigun10}. Here we consider the comparisons in both systems of reference, by comparing galaxy colours in the observed frame, and the galaxy mass function using data-modelled stellar masses.

Finally, we compare the light-coned BOSS mass function with mass functions from the literature.

The paper is organised as follows. In Section 2 we introduce the BOSS data, in Section 3 we detail the stellar mass calculation and in Section 4 we present and discuss the results relative to the stellar masses of BOSS galaxies. In Section 5 we perform the comparison with semi-analytic models and in Section 6 we summarise the work and draw conclusions. 

Throughout the paper the cosmology from WMAP1, i.e. $\Omega_{M}=0.25$, $H_{0}=0.73~\rm {km~s^{-1} Mpc^{-1}}$, $\Omega_{T}=1$, is assumed for consistency with the galaxy evolution models \citep{guoetal11,henetal12}\footnote{Note that for the DR9 release, see Section 2, a slightly different cosmology has been adopted, namely 
$\Omega=0.258$, $H_{0}=71.9~\rm {km~s^{-1} Mpc^{-1}}$, $\Omega_{T}=1$,. We checked that this implies a negligible effect on stellar masses.}.
\section{BOSS galaxy data: the 'constant mass' sample definition}
\label{eps}
The BOSS survey \citep{schwhieis10} aims at constraining the late time acceleration in the Universe via Baryon Acoustic Oscillations (Eisenstein et al. 2005; see also Anderson et al. 2012 for the first results on BOSS)\nocite{eisetal05,andetal12}, with an observational effort of galaxy spectroscopy and photometry over five years, that started in Fall 2009. An overview of BOSS is given in \cite{dawetal13}. Below we summarise the key aspects that are relevant to this paper.
BOSS is one of four surveys of the SDSS-III collaboration \citep{eisetal11} using an upgrade of the multi-object spectrograph (Smee et al. 2012, {\it submitted}) on the 2.5m SDSS telescope \citep{gunetal06} located at  Apache Point Observatory in New Mexico.
BOSS obtains medium resolution ($R=2000$) spectra for galaxies, QSOs and stars in the wavelength range $3750-10000$~\AA. Standard SDSS imaging using a drift-scanning mosaic CCD camera \citep{gunetal98} is obtained for luminous galaxies over the redshift range 0.3 to 0.7, selected to be the most massive and with a uniform mass sampling with redshift \citep{whietal11,eisetal11}. The acquired photometry has been released with the Data Release 8 \citep[DR8,][]{aihetal11}, and the first set of spectra will be made publicly available with the Data Release 9 (DR9), in Summer 2012 (Ahn et al. 2012).

For the project, we calculated photometric stellar masses for BOSS galaxies. We use the galaxy spectroscopic redshift determined by the BOSS pipeline (Bolton et al. 2012; Schlegel et al. 2013, {\it in prep.}) and standard $u,g,r,i,z$~SDSS photometry \citep{fuketal96} for performing spectral energy distribution (SED) fitting at fixed spectroscopic redshift in order to obtain a best-fit model and from it an estimate of the stellar mass (see Section~3). The values of stellar mass and the routines to perform the same calculations for the rest of the BOSS survey will be made publicly available through DR9 in Summer 2012.\footnote{For this work we selected objects with solid spectroscopic redshift determination (corresponding to the flag {\it zwarning}=0) and we used the primary spectroscopic observation available (using flag {\it specprimary=1}). These flags select a total number of galaxies which is slightly lower than what will be available with DR9.}

The BOSS galaxy sample consists of two parts. The high-redshift or CMASS (i.e. constant mass) sample, mostly containing galaxies with a redshift of 0.4 or larger and aimed at defining a galaxy sample with homogeneous stellar mass; a lower-redshift sample (LOWZ), which is included in BOSS in order to increase the effective area and to allow for comparison with the SDSS I \& II samples.

The {\it constant mass} CMASS selection is achieved by tracking the location in observed-frame colours and magnitudes, of model galaxies of different mass as a function of redshift. A passively evolving model (Maraston et al. 2009) is adopted. The method was checked on a sample of galaxies from the AGES survey \citep{kocetal12}, by deriving their stellar masses via broad-band $u,g,r,i,z$ SED fit as in this paper. 

\begin{figure}
  \includegraphics[width=0.49\textwidth]{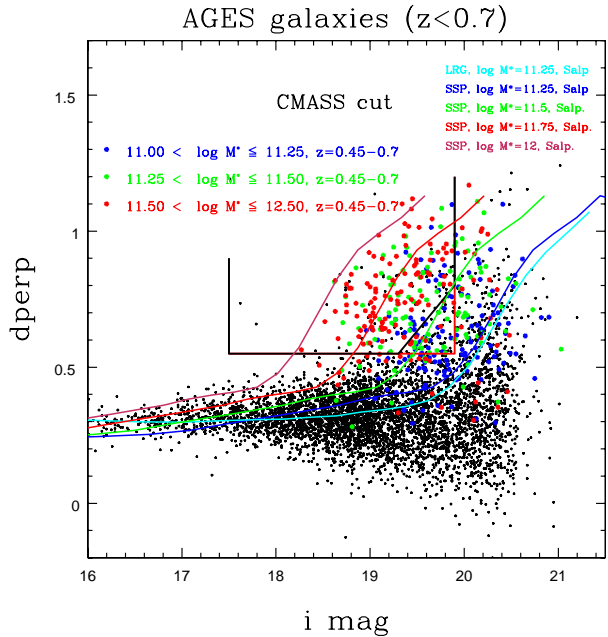}
  \caption{$d_{\perp}=((r-i)-(g-r))/8$ vs $i$ observed-frame colour-magnitude diagram of galaxies from the AGES survey (Kochanek et al. 2012), coloured-labeled by stellar mass (detailed in the panel). Black points indicate galaxies with $M^{*} < 10^{11} M_{\odot}$. Solid lines highlight the target selection, which further picks galaxies lying at a redshift larger than 0.4.}
  \label{fig:agescmass}
\end{figure}

Figure~\ref{fig:agescmass} displays the location of AGES galaxies with different stellar masses (plotted in different colours) on the target selection plane of observed $i$~magnitude vs the composite colour $d_{\perp}(=(r-i)-(g-r)/8.0)$. Coloured points indicate $M^{*}\gapprox 10^{11} M_{\odot}$, black points galaxies with a lower mass. In addition to the mass selection, the redshift selection is based on the $r-i$~colour, which traces the D-4000 \AA\ break in galaxy spectra in this redshift range \citep{eisetal01}. The final mass and redshift selection is achieved through a {\it sloping} cut, corresponding to the redshift evolution of models with various total stellar masses (solid lines in Figure~\ref{fig:agescmass}). The colour equations for the target selection write as:  $17.5 < i < 19.9$, $d_{\perp} > 0.55$, and $i < 19.86+1.6\cdot d_{\perp}-0.8)$ where i is the {\it cmodel} (see below) magnitude,  for CMASS; $16 < r < 19.5$, $r < 13.6 + {c_{\parallel}/0.3}$, and ${c_{\perp}} < 0.2$ where r is the {\it cmodel} (see below) magnitude, for LOWZ (see Eisenstein et al. (2011) and Dawson et al. (2013) for further details).\footnote{The composite colours $c_{\parallel}$, $c_{\perp}$ are defined as  $c_{\parallel} = 0.7(g-r) + 1.2(r-i-0.18)$, $c_{\perp}=(r-i)-(g-r)/4-0.18$.}

\begin{figure}
  \includegraphics[width=0.49\textwidth]{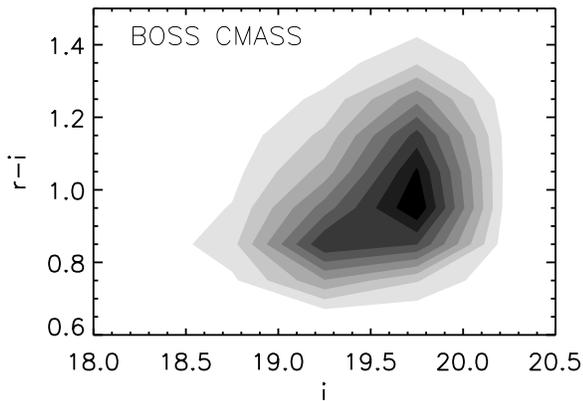}
  \caption{Observed-frame colour ($r-i$) - magnitude ($i$-model) diagram of BOSS galaxies in the high-redshift CMASS sample. Magnitudes are galactic extinction corrected (see text).}
  \label{fig:cmcmass}
\end{figure}

Figure~\ref{fig:cmcmass} shows as visualisation the actual CMASS sample of BOSS galaxies in an observed-frame colour-magnitude diagram. The effectiveness of CMASS at selecting a constant stellar mass will be quantified and discussed in Section 4.


The BOSS data sample, including both CMASS and LOWZ, that was acquired through September 2011, contains over 400,000 galaxies\footnote{We additionally calculated the stellar masses of DR7 galaxies with the same method, which will be published separately}. In this paper we focus on the CMASS $z\gapprox0.4$~sample for the comparison with galaxy formation models. 

Spectroscopic redshifts are determined from BOSS spectra using the latest version of the SDSS Spec 1D pipeline and an extensive set of templates, based on both stellar empirical spectra as well as population models (Bolton et al. 2012; see also Schlegel et al 2013, {\it in prep.}, which explain the procedure to obtain spectra which are input to the pipeline). The redshift success for CMASS has the impressive figure of $\sim98\%$~(Anderson et al. 2012, Table 1) and is even better for the low-redshift (LOWZ) sample.
\
Different magnitude definitions are available for galaxy photometry in SDSS. {\it Model} magnitudes aim at providing accurate colour information, whereas {\it cmodel} magnitudes are better for accurate total luminosity \citep{stoetal02}.\footnote{http://www.sdss.org/dr7/algorithms/photometry.html\#magmodel} For an SED fit aimed at mass determination we need both types of accuracy, so we decided to use {\it model} magnitudes  but scale the values using {\it cmodel} magnitudes in the $i$-band. This scaling results in a constant shift of the entire SED. We choose the $i$-band as this maps into $r$-rest-frame at the BOSS redshifts, which is the base for model magnitudes. We have performed separate SED-fit calculations using either {\it model} or {\it cmodel} magnitudes and find that this choice mostly affects the scatter. Finally, we applied extinction correction for Milky Way reddening using \citet{schetal98} values.
 It should be noted that this method of combining magnitudes is the official method adopted for the galaxy target selection for BOSS. 

Typical photometric errors of {\it model} magnitudes are $1.00, 0.17, 0.06, 0.04, 0.09$ in $u,g,r,i,z$, respectively. These are averages evaluated on 331,915 BOSS galaxies at redshift $\sim~0.55$.  Also errors are scaled to {\it cmodel} magnitudes\footnote{The scaling writes as $magscaled_{err[ugriz]}=magscaled[ugriz]\cdot modelmag_{err[ugriz]}/modelmag[ugriz]$}, in order to preserve the S/N. 

\section{Stellar mass calculation}
\label{sec:masscalculation}
Photometric stellar masses ($M^{*}$) are obtained with the standard method of SED fitting \citep[e.g.][]{sawyee98}, where observed magnitudes are fitted to model templates to obtain a model stellar population that best matches the data. The normalisation of this model to the data provides an estimate of the galaxy stellar mass. 

The fitting can be performed at fixed redshift or by leaving the redshift as a free parameter to be adjusted and determined with the fitting method itself. Here - by virtue of the BOSS spectroscopic redshift - we can use the fixed redshift option. The adopted fitting method and stellar templates are described below.
\subsection{Galaxy model templates}
We adopt two sets of templates in order to encompass plausible variations in the star formation histories of BOSS galaxies. 

First is a passive template, which we found to best match the redshift evolution of luminous red galaxies (LRGs) from the 2dF SDSS LRG and Quasar (2SLAQ) survey (Cannon et al. 2006) up to a redshift of 0.6 \citep{cooetal08,CMetal09a}. This passive model is the superposition of two single-burst models with identical age and very different metallicity, namely solar and 0.05 solar, in proportion as 97\% and 3\% by mass. Age is the only parameter of this model. The base model is the Maraston \& Str\"omback (2011) model based on the \citet{pic98} empirical stellar library. The reason for the better match, with respect to standard solar metallicity passive models or models with star formation \citep[e.g.,][]{eisetal01,waketal06} is twofold. 
First, we use empirical model atmospheres in place of the standard Kurucz-type ones, which produce a slightly "bluer" $g-r$ and a slightly "redder" $r-i$ as the galaxy data suggested. This effect, though not associated with a choice of star formation, is important at the end of improved modelling. The effect of various model atmospheres/empirical stellar libraries on the optical spectral shape of a stellar population model is discussed in detail in \citet{marstr11} where the same spectral shape as in empirical libraries is found in the new-generation theoretical model atmospheres calculated with the software MARCS \citep{gusetal08}. The correct shape of the model around the $V$-band has been confirmed using data of star clusters in M31 \citep{peaetal11} as well as in the Milky Way \citep{marstr11}.

Second, metal-poor stars add blue~light to the passive metal-rich model which, opposite to young stars, is slowly evolving with redshift, in better agreement with those data. This two-component model can be explained as to represent a metal-poor halo in these massive galaxies. 

In addition to the passive model, we consider a suite of templates with star formation, namely exponentially-declining star formation $e^{-t/\tau}$, with $\tau=0.1,0.3,1$ Gyr and "truncated" models, where star formation is constant for a certain time elapsing from the beginning of star formation, which we call 'truncation time', and zero afterwards. Here we used truncation times of 0.1, 0.3 and 1 Gyr.

Each star formation history is composed of 221 ages, and is calculated for four different metallicities, namely $0.2, 0.5, 1$ and $2$~solar. This selection of templates was used in \citet{dadetal05} and \citet{CMetal06} for the SED-fit of passive galaxies at $z\sim2$. We refer to this second template as SF.
\begin{figure*}
  \includegraphics[width=\textwidth]{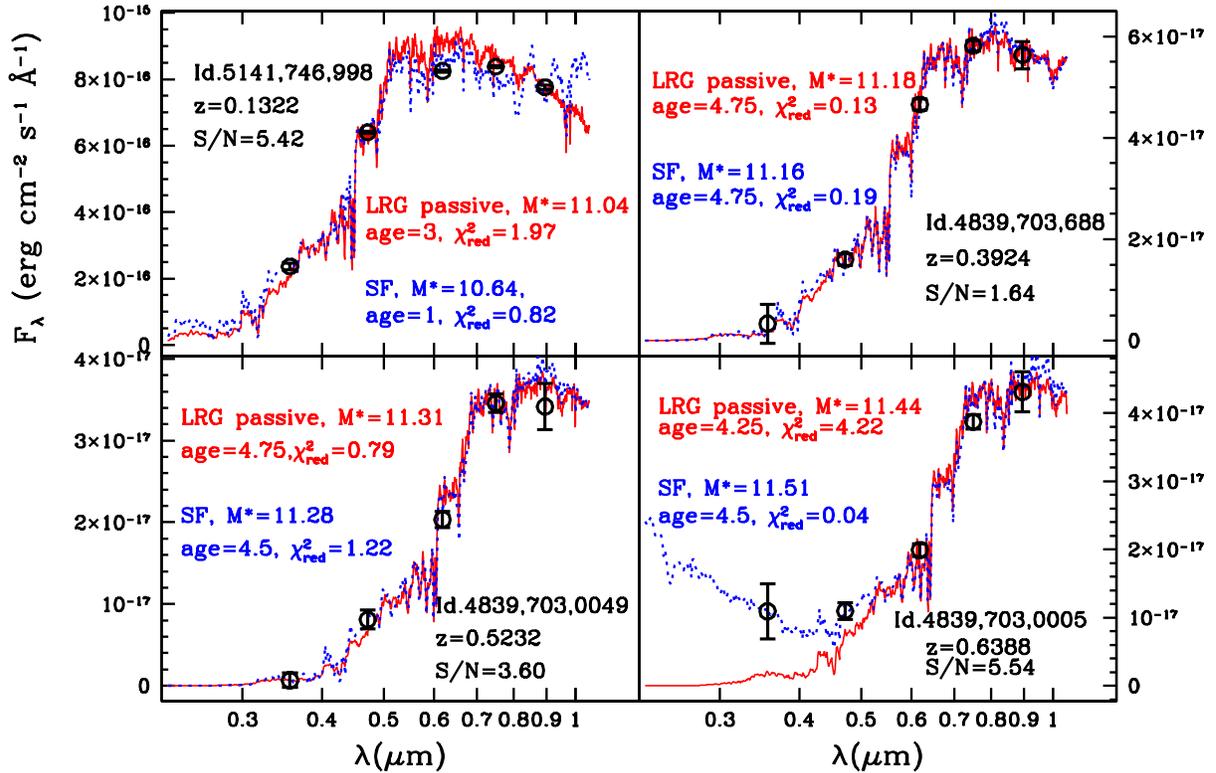}
  \caption{Examples of SED-fit results for four BOSS galaxies, in order of increasing spectroscopic redshift from top left to bottom right. Red and blue lines display the best-fit models and labels show $\log \rm M^{*}/M_{\odot}$, age (Gyr), reduced $\chi^2$, as obtained using the LRG-passive and the star forming (SF) templates, respectively. Object Id. (given as plate, mjd, fiberid), spectroscopic redshift and median photometric S/N are indicated.}
  \label{fig:sedfit}
\end{figure*}
Both template models were calculated for a \citet{sal55} and a \citet{kro01} initial mass function (IMF), and in both cases the stellar mass lost due to stellar evolution is subtracted from the total mass budget. The stellar mass budget including white dwarf, neutron star and black-hole remnants follows our previous calculations \citep{CM98,CM05} and is based on the initial mass versus final mass~relations by \citet{rencio93}. For a single burst population following passive evolution, the fraction of mass lost is around $30$~to $40\%$~depending on the assumed IMF (Maraston 2005, Figure 27).\footnote{As stellar mass losses are not always subtracted from the total mass in the literature, we provide values with and without the inclusion of this effect.}

\subsection{Fitting code and method}
\label{fitting}
We employ the fitting code {\it HyperZ} \citep{boletal00}, and in particular an adapted version of it, named {\it HyperZspec}, in which the SED fitting is performed at a fixed spectroscopic redshift. This latest version also uses a finer age grid of 221 ages for each star formation history, instead of the 51 adopted in earlier versions\footnote{The latest version of the {\it HyperZspec} code was kindly made available to us by Micol Bolzonella.}. The use of a denser grid, though not changing any result appreciably, allows for a better recovery of galaxy properties \citep{pfoetal12}. The code can be used with various stellar population models \citep[see][]{boletal10,CMetal06,CMetal10}. For this work we adopt the models described in Section 3.1.

The fitting procedure is based on maximum-likelihood algorithms and the goodness of the fit is quantified via reduced $\chi^{2}$ ($\chiquadr$) statistics. The code computes $\chiquadr$ for a large number of templates, which differ in their SFHs, and identifies the best-fitting template. It should be noted that in the reduced $\chiquadr$~calculated via HyperZ, the degrees of freedom are only set by the number of photometric filters (minus unity), and not by the actual intrinsic degree of freedom of the adopted template (e.g. age, metallicity, star formation history, reddening). This implies that the $\chiquadr$~obtained with different templates should not be compared quantitatively. The code does not interpolate on the template grids, hence the template set must be densely populated. The internal reddening $E(B-V)$~as parametrized by various laws can be used as an additional free parameter. 

An important feature of our analysis is that we do not include reddening in our fitting procedure. This is because our study of the SED fit of simulated galaxies (Pforr et al. 2012) shows that the level of degeneracy increases and solutions with unlikely low ages and substantial dust may have favourable $\chiquadr$ values when reddening is included as a free parameter. This problem is known as age/dust degeneracy (e.g. Renzini 2006 for a review). These young, dusty models provide a good representation of the photometric SED, but the derived mass significantly underestimates the true total galaxy mass (Pforr et al. 2012, Figure~11). Our work further shows that this effect is more severe in old galaxies that have experienced a recent, small burst of star formation. Such galaxies are, in the simulations and likely in the real Universe, mostly found at redshift below 1, i.e. in the realm of BOSS observations. Higher-redshift galaxies - by having overall younger stellar populations and a smaller spread in age - suffer less from these degeneracies.\footnote{Note that these results may depend on the set of adopted templates in terms of star formation history, and on the intrinsic amount of dust. The mocks we use have a limited amount of dust, up to $E(B-V) =0.3$ at $z\sim3$ and lower at lower redshifts. Hence, we cannot extrapolate these finding to highly dusty galaxies for which most of the population is newly born and reddened. Here the non inclusion of reddening may lead to overestimate the age, hence the mass, and would provide bad fits which could be hard to select. We may address these cases in future work.}

In summary, to keep our SED-fit mass estimates as protected as possible from the age-dust degeneracy, we do not include reddening. Reddening for BOSS galaxies can be quantified through emission-line studies (Thomas et al. 2013, Figure 8) and is included in galaxy spectral fitting by Chen et al. (2012) and Toieiro et al. (2012). None of these works find the bulk of BOSS CMASS galaxies to be dusty, as they are selected to be mostly quiescent. For example, from the emission lines we get an average reddening of $E(B-V)\sim0.05$ (Thomas et al. 2013). This value is also consistent with the observed morphologies of the sample of BOSS galaxies we could cross-match to COSMOS, where we find that $\sim73\%$~of BOSS galaxies are early-types (Masters et al. 2011, see Section~4.1)\nocite{masetal11}. 

The age of the best-fits is the age at the onset of star formation in that model, hence it corresponds to the formation age. Fitted ages are constrained to be younger than the age of the Universe in the adopted cosmology. We also apply age cutoffs to the templates. 
The minimum allowed fitting age for the passive LRG model is 3 Gyr. This corresponds to the assumption that the descendants of these galaxies are 10 Gyr old at redshift zero, and in our adopted cosmology the look-back time to redshift 0.8 (roughly corresponding to the maximum redshift sampled in  BOSS) is $\sim~7$~Gyr. The set of a minimum age in the fitting minimises the probability of underestimating the stellar mass by obtaining too low an age. This will be shown and discuss in Section 4.1.  Should we relax this prior, we would obtain a fraction of galaxies amounting to 20-30\% depending on redshift which would have somewhat lower ages, hence lower masses. However, as we shall see in Section~4.1, the minimum age of 3 Gyr seems to guarantee the best mass recovery, hence we shall retain this prior. We have also verified that the effect on the resulting mass function is marginal (see Figure \ref{fig:massfuncnotmin}, Appendix C).

On a similar ground, we apply an age cutoff to the star forming model of 0.1 Gyr, which is typically assumed in SED-fit of star-forming galaxies (e.g. Bolzonella et al. 2010; Maraston et al. 2010). An age cutoff of this size helps minimising the event of fitting for too low ages. However, we have further verified that an additional correction to the stellar masses for star forming galaxies is required, which we shall discuss in Section 4.1. Finally, the mass is calculated with a routine developed in \citet{dadetal05} and \citet{CMetal06}, and extended  for this project for properly handling large databases.

\begin{figure}
  \includegraphics[width=0.49\textwidth]{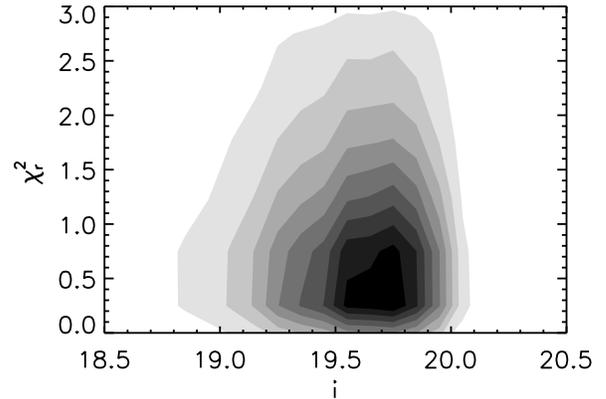}
  \caption{Reduced $\chi^2$ (\chiquadr)~as a function of observed-frame $i$-model magnitude for the SED fits of BOSS galaxies.}
  \label{fig:chi}
\end{figure}

A few examples of SED-fits are shown in Figure~\ref{fig:sedfit}, for randomly chosen galaxies at various redshifts. The best fit population parameters obtained using the two templates - the passive LRG and the suite with star formation (SF) - are indicated in red and blue, respectively. BOSS data are shown as circles. Excellent fits are obtained, in general with both templates, even for objects with low S/N\footnote{The S/N values plotted in the figure are photometric, but we have verified that the same can be concluded when one uses the spectral S/N.}. 
The distributions of reduced-$\chi^2$~is shown in Figure~\ref{fig:chi} as a function of the $i$-model magnitude in observed-frame. The \chiquadr\ values do not depend on the object's magnitude, and we have checked they also do not depend on the object's redshift.
\begin{figure*}
  \includegraphics[width=0.8\textwidth]{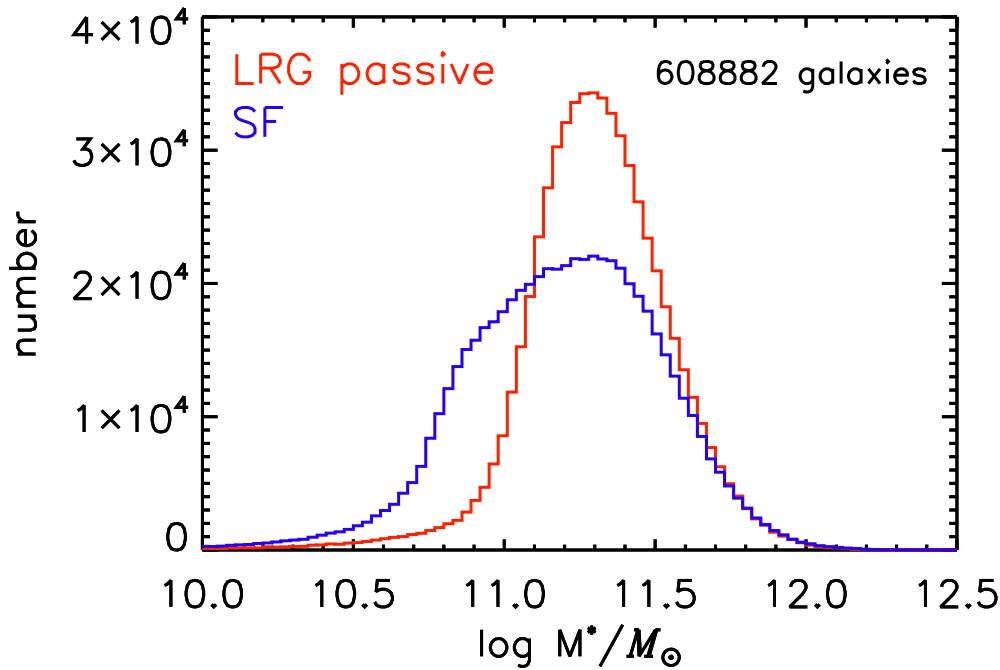}
  \caption{Photometric stellar masses of BOSS galaxies in the first two years of data. The two histograms show $\log \rm M^{*}/M_{\odot}$~as obtained with different galaxy templates: the LRG passive model of \citet{CMetal09a} (red), in which a small fraction ($3\%$) of old metal-poor stars is added to a dominant metal-rich ($Z=Z_{\odot}$) population, both being coeval and in passive evolution, and a set of templates with star formation (blue), ranging from $\tau$-models to constant SF. Stellar masses obtained with the SF template are systematically lower due to the lower M/L~of young populations. Calculations shown here refer to a Kroupa IMF and included mass-losses from stellar evolution. Average errors on $\log \rm M^{*}/M_{\odot}$~are 0.1 dex (cfr. Figure~\ref{fig:masserrors}).}
  \label{fig:masstemp}
\end{figure*}

The fitting procedure gives the best-fit model corresponding to the minimum $\chi^2$~and the probability distribution function (PDF) of neighbouring solutions for different cuts in $\chi^2$~above the minimum. Interestingly, we find that the difference in stellar mass between the best-fit value and the median PDF value is only 0.03 dex in case of the LRG template, and at most 0.1 dex in case of the templates with star formation, due to the higher number of neighbouring solutions with similar $\chi^2$.
\section{Results}
\label{sec:results}
We have calculated the photometric stellar masses $M^{*}$ for $\sim 400,000$ massive luminous galaxies from the first two years of data of the SDSS-III BOSS survey. The calculations of stellar mass is released with the Data Release 9 (DR9), as well as ages, star formation histories (SFH), star formation rate (SFR), and metallicities, for each of the two template fittings and the two adopted IMFs. Ages, SFRs and stellar masses are provided with their 68\% confidence levels. We also derive median stellar masses by taking the median of the PDF and list them together with their 68\% confidence levels. In each case, we provide $M^{*}$~with and without stellar mass-loss due to stellar evolution. We note here that, even if we provide all quantities derived through the SED-fit, the procedure is studied as to maximise the quality of $M^{*}$~determination. The other by-products of the fits should be considered less robust. For example, as we do not include reddening from dust, the age of the most recent burst maybe ill determined. Also, metallicity does not vary in the templates. Future work will be invested in a more detailed spectral analysis.

Figure~\ref{fig:masstemp} shows the distribution of stellar masses of BOSS galaxies for the combined CMASS and LOWZ samples, for the LRG (red) and the SF template (blue). Plotted values refer to the Kroupa IMF, and stellar mass loss has been accounted for in the calculations. For the results obtained with the LRG template, the mass histogram is thin and well defined, pointing to a uniform mass distribution as a function of redshift as was the aim of the BOSS target selection (White et al. 2011; Eisenstein et al. 2011; Dawson et al. 2013). We quantify this later in the section.

\begin{figure}
  \includegraphics[width=0.49\textwidth]{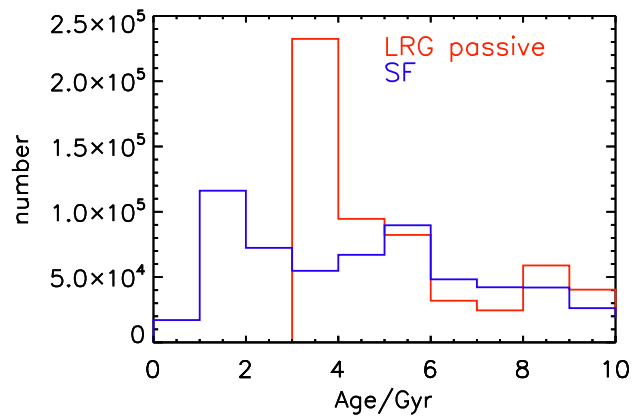}
  \caption{The distribution of stellar ages obtained for BOSS galaxies using different templates for SED fitting, namely the LRG passive template (red) and the template with star formation (blue).}
  \label{fig:ages}
\end{figure}

\begin{figure}
  \includegraphics[width=0.49\textwidth]{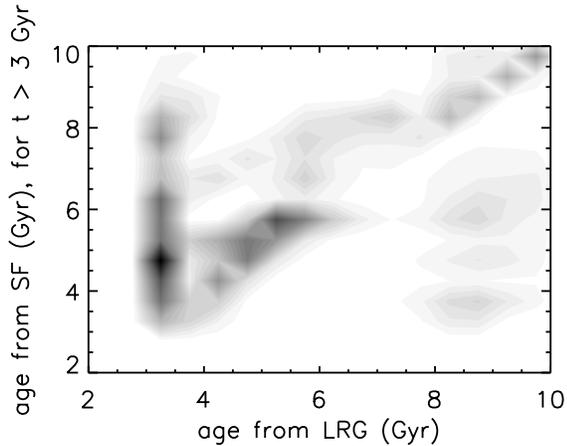}
  \caption{Comparison of ages of individual galaxies, for ages larger than 3 Gyr, obtained with the SF template versus those from the LRG template. The fraction of galaxies with correlated ages (age difference within 0.5 Gyr) is $\sim 25\%$.}
  \label{fig:agevsage}
\end{figure}

The results for both templates agree reasonably well in indicating a peak stellar mass of $\sim 11.3~\rm \log M$~(for a Kroupa IMF, $1.6$~higher for a Salpeter IMF). Stellar masses derived with the SF template (blue) show an excess of lower mass values which is due to the lower ages for some of the galaxies derived with this template, see Figure~\ref{fig:ages}. 
Except for this, the age distributions agree remarkably well for ages larger than 3-4 Gyr, independently of the adopted template, which confirms the homogeneous nature of the CMASS sample \citep[see also ][]{tojetal12}. Note that the ages of individual galaxies do not necessarily agree, as shown in Figure~\ref{fig:agevsage}, where we plot ages from the SF template (for values higher than 3 Gyr) vs ages from the LRG template. Ages obtained with the SF template are older by $\sim2$ Gyr with respect to those from the LRG template. This happens because the SF template allows for extended star formation hence the age (which is the time elapsed since the beginning of star formation) obtained with this template can be larger and able to fit the same set of data. In spite of these differences for a fraction of galaxies, individual masses agree well due to compensating effects between age and star formation history, Figure~\ref{fig:massvsmass}.

In Appendix A we discuss in detail the comparison with other stellar mass calculations performed in BOSS, while in Appendix B we present rest-frame magnitudes that are a by-product of the fitting and will be available via DR9.

\begin{figure}
  \includegraphics[width=0.49\textwidth]{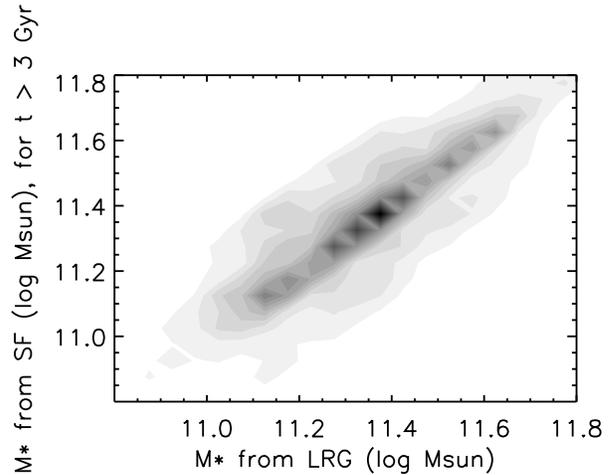}
  \caption{Comparison of stellar masses of individual galaxies, for ages larger than 3 Gyr, obtained with the SF template versus those from the LRG template. The scatter in the correlation is $\sim 0.13$ dex.}
  \label{fig:massvsmass}
\end{figure}
\begin{figure*}
  \includegraphics[width=0.8\textwidth]{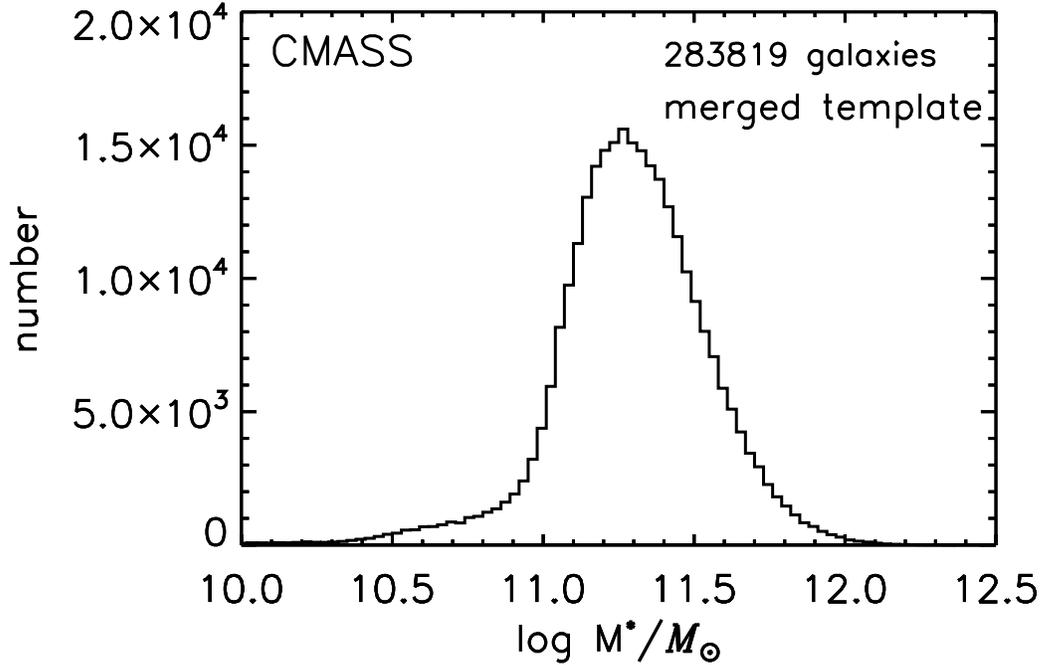}
  \caption{The final $M^{*}$~distribution of BOSS/CMASS galaxies where values of stellar mass obtained with different templates are assigned according to the galaxy type - passive or star-forming - using the cut in apparent colour $g-i\sim~2.35$. Galaxies on the red side of the colour cut get $M^{*}$~from the passive LRG template and those on the  blue side from the SF template. The total stellar mass distribution of BOSS galaxies peaks at $\sim~11.3~M_{\odot}$, for a Kroupa IMF, with a mean of $\sim~11.27~M_{\odot}$ and a FWHM of $\sim 0.7$~dex.}
  \label{fig:massmerged}
\end{figure*}

As mentioned in Section 2 and shown in Figure 1 the target selection for the BOSS survey aimed for a uniform mass sampling as a function of redshift. We can now test whether this goal has been achieved.  
Figures~\ref{fig:masszdr9lrg} show the stellar mass distributions in various redshift bins, for the calculations referred to the two different templates, LRG passive and SF, for the combined CMASS plus LOWZ samples.

A remarkably uniform mass sampling is achieved in a large redshift range spanning between redshift 0.2 and 0.6, when stellar masses are determined with the LRG passive template.\footnote{The mean $\log \rm M^{*}/M_{\odot}$ (for a Kroupa IMF, including stellar mass losses) in the various redshift bins are: for the LRG template,  11.33 at $0.2 \lapprox z\lapprox 0.4$, 11.27 at $0.4 \lapprox z\lapprox 0.5$, 11.26 at $0.5 \lapprox z\lapprox 0.6$, 11.41 at $0.6 \lapprox z\lapprox 0.7$, and 11.61 at $z\gapprox 0.7$; for the SF template,
11.2028 at $0.2 \lapprox z\lapprox 0.4$, 11.19 at $0.4 \lapprox z\lapprox 0.5$, 11.14 at $0.5 \lapprox z\lapprox 0.6$, 11.26 at $0.6 \lapprox z\lapprox 0.7$, and 11.31 at $z\gapprox 0.7$; for the merged template, 11.31 at $0.4 \lapprox z\lapprox 0.5$, 11.32 at $0.5 \lapprox z\lapprox 0.6$, 11.41 at $0.6 \lapprox z\lapprox 0.7$, and 11.53 at $z\gapprox 0.7$.} At $z\gapprox 0.6$, the mass distribution is skewed towards higher values, which is probably due to the magnitude limit of the survey. From these plots we infer that BOSS becomes incomplete at $z\gapprox 0.6$ and $\log \rm M^{*}/M_{\odot}\lapprox 11.3$. This suggestion will be qualitatively confirmed when we will compare the BOSS mass function with literature values (Section 5.2.1).

The assumed template impacts the uniformity of the mass sampling, as should be expected. Figure~\ref{fig:masszdr9lrg} (lower panel) shows that, when interpreted with templates including star formation, a fraction of BOSS galaxies get lower stellar masses, which leads to secondary peaks in the mass distributions. The highest redshift bin is the most strongly affected by the
template choice because galaxies get younger at higher redshift and the larger age spread allowed by the SF template emphasises age hence mass difference.

We also note that the mass distribution in the lowest redshift bin (the LOWZ sample) is narrower than those  at higher redshift, in particular for the passive template fit. This happens because the LOWZ sample has a narrower colour selection hence is populated by more uniform galaxies with respect to CMASS, which extends to a bluer colour selection. This colour span can be appreciated in Figure 9 of Thomas et al. (2013). 
In addition, the LOWZ sample is at lower redshift hence contains more evolved galaxies. We shall take this into account when deciding upon the most suitable template in the next section.

\subsection{The final BOSS mass distribution: sorting templates by galaxy colours}
As described in the previous section, we calculate stellar masses with two templates in separate runs. Hence, each BOSS galaxy has two possible values of $M^{*}$. This will be useful when the stellar masses of BOSS galaxies are used for comparison with results from other surveys in which various templates are adopted. Nonetheless, for most science applications it would also be useful to have one preferred choice of $M^{*}$. 

In this section we describe a colour criterion to assign stellar mass values from different templates to observed galaxies which is based on the galaxy colour. Furthermore, as we will show, the selection in colour is very close to a selection in morphology, with early-type galaxies being almost always on the red side, and star forming galaxies to the blue. 

In Masters et al. (2011), we cross-matched the BOSS sample with the COSMOS survey \citep{capetal07} which provides high resolution $I$-band imaging from the Hubble Space Telescope (HST) over 2 square degrees. The cross-match yields 240 BOSS target galaxies for which detailed morphological information was obtained. Visual inspection of the COSMOS images was used to select early and late-type galaxies under the typical classification scheme. Any smooth galaxy was determined to be early-type, including those with the appearance of a smooth disc (S0 or lenticular). To be called late-type the galaxies needed to have visible spiral arms or be obviously edge-on discs. Edge-on discs might be confused between S0 or spiral, and have been marked separately on Figure 4 of Masters et al. (2011). \footnote{All images can be inspected at http://www.icg.port.ac.uk/~mastersk/BOSSmorphologies/.}Ê

We found that $\sim73\%$~of the galaxies in CMASS are early-types, and the rest $\sim27\%$~is composed by late-types. 
Critical to the analysis of the present paper, we defined a simple colour criterion of $g-i$, namely $g-i\gapprox~2.35$, which allows us to separate early-types from later-types with better than 90\% purity. Here we employ this colour criterion to assign mass values obtained with different templates to the different morphological classes. We use the best fit LRG mass for objects with $g-i\gapprox~2.35$,  and the best fit SF mass for galaxies with $g-i\lapprox~2.35$, which is the location of most spirals. It should be noted that the fraction of CMASS galaxies with $g-i\lapprox~2.35$ is only 30\% in the full COSMOS subsample, and the fraction of early-types among these is only 20\%, so clearly a minority. This demonstrates the strong (and well known) correlation between morphology and colour, with early-type galaxies being almost always redder than late-type galaxies.

The final total $M^{*}$~distribution of BOSS CMASS galaxies is shown in Figure~\ref{fig:massmerged}. Similar to Figure~\ref{fig:masstemp}, the total mass distribution still peaks at $\log \rm M^{*}/M_{\odot} \sim11.3$ (for a Kroupa IMF) and is dominated by the mass values obtained with the LRG template, as the majority of galaxies in CMASS is of early-type. The adoption of the values obtained with the SF template implies an excess of galaxies with $\log \rm M^{*}/M_{\odot} \sim10.8$ with respect to the distribution obtained using the LRG template.

The scatter around best-fit masses for individual galaxies, expressed as $\log(M//M_{\rm best-fit}$), is shown in Figure~\ref{fig:masserrors}. The scatter is $\sim0.1$~dex, it is approximatively symmetric and we have verified that is independent of galaxy mass.

We have also tested the goodness of our template choice with mock galaxies with known input mass. We use galaxies from a semi-analytic model (Tonini et al. 2009, which is based on the GALICS semi-analytic model by Hatton et al. 2003), picked out of the full merger tree to be representative of the range in mass and star formation rate predicted by the models. In practice, the star formation rate in the mocks can be very low, but it is never zero, strictly speaking (cfr. Figure 4 in Pforr et al. 2012). These mock galaxies coincide with the mock star forming option as used in Pforr et al. (2012). We treat the mocks as observed galaxies and calculate their stellar mass via SED fitting, which we then compare to their actual stellar mass.

Figure~\ref{fig:mockLRG} shows the results for mocks at redshift 0.5, where input stellar masses ($x$-axis) are compared to photometric stellar masses obtained via SED-fit to broadband $u,g,r,i,z$~photometry with the LRG passive template ($y$-axis). The red colour highlights those mocks that have $g-i\gapprox2.35$, which corresponds to the colour region where we use the LRG template in the BOSS sample. 
The various panels show the results having applied different minimum age cutoffs to the fitting procedure, increasing from 0.1 to 3 Gyr  from top left to bottom right. As can be seen, the mass offset between intrinsic and recovered mass decreases at increasing minimum age cutoff reaching a minimum at 3 Gyr. Here the stellar masses of these "reddest" galaxies are well recovered with the LRG template, with a scatter consistent with zero (further visualised as an histogram in Figure~\ref{fig:histodmLRG}).\footnote{The 'red' mocks display a constant offset with mass in Figure 12. This is because their SFR does not change much as a function of mass.   
Note that this is not what happens in the real BOSS galaxies, for which only a fraction of the red galaxies gets a younger age (hence a lower mass) when the age cutoff is relaxed, as is discussed in Appendix C. This is a difference between semi-analytic and real galaxies, the star formation properties of which are probably a much stronger function of stellar mass, as we know from results on downsizing, e.g. Thomas et al. (2010).} Black points represent the results for mock galaxies with bluer colours, $g-i\lapprox2.35$. For these, the application of the LRG passive template would lead to an overestimate of the mass, so for this type of objects in BOSS we use stellar masses obtained with the SF template. 

\begin{figure}
  \includegraphics[width=0.49\textwidth]{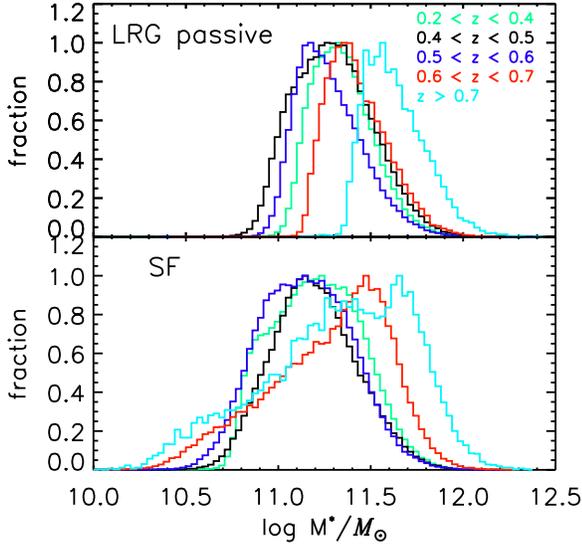}
  \\
  \\
  \caption{The distribution of stellar mass in the combined CMASS and LOWZ sample, in various redshift bins (normalised to the peak mass value in each bin), for results obtained with the LRG passive template (upper panel) and the SF template (lower panel). The mass distribution is fairly uniform in the redshift range $0.2\lapprox z \lapprox 0.6$ (cfr. green, black and blue histograms).}
  \label{fig:masszdr9lrg}
\end{figure}
\begin{figure}
 \includegraphics[width=0.49\textwidth]{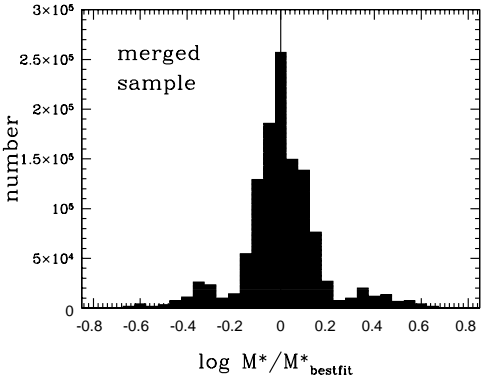}
  \caption{Scatter around best-fit masses, expressed as $\log(M//M_{\rm best-fit}$), of galaxies for fits from the merged-template sample. The scatter is calculated form individual PDFs as the 68\% confidence interval with respect to the best-fit $M^{*}$ solution. On average, the scatter is [$-0.1,+0.1$] dex with respect to the best-fit value.}
 \label{fig:masserrors}
\end{figure}
\begin{figure}
  \includegraphics[width=0.49\textwidth]{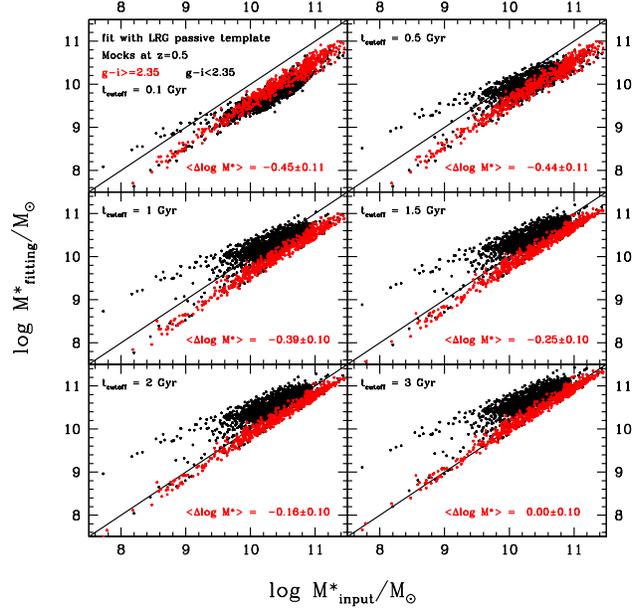}
  \caption{The recovery of stellar mass via SED fitting with the passive LRG template as a function of the minimum age cutoff assumed in the fitting procedure. The mass obtained via SED fitting ($y$-axis) of mock galaxies from semi-analytic models (Tonini et al. 2009) at redshift 0.5 is compared to their intrinsic mass ($x$-axis), for several age cut off from 0.1 to 2.5 Gyr. The mass discrepancy for red galaxies 
 ($g-i\gapprox 2.35$, red points) decreases as a function of the age cutoff, reaching optimal values for $t\sim3$~Gyr, where stellar masses are recovered with a scatter of only 0.06 dex.}
  \label{fig:mockLRG}
\end{figure}
\begin{figure}
  \includegraphics[width=0.49\textwidth]{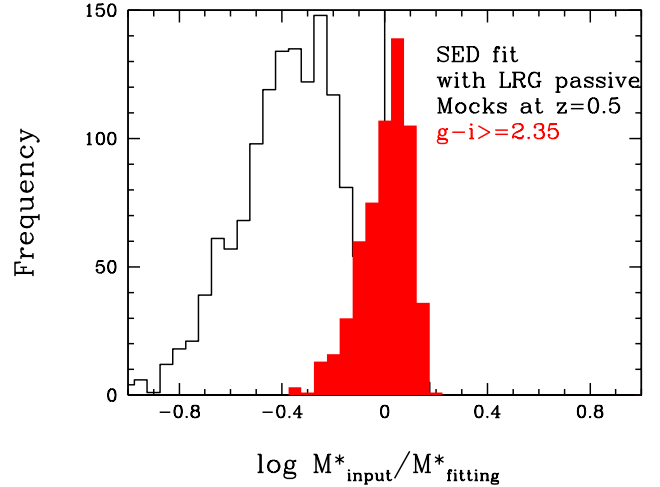}
  \caption{Histogram of mass difference between intrinsic mass and recovered mass, for the LRG template with age cutoff of 3 Gyr applied to red galaxies (cf. Figure~\ref{fig:mockLRG}, bottom, right panel).}
  \label{fig:histodmLRG}
\end{figure}
\begin{figure}
  \includegraphics[width=0.49\textwidth]{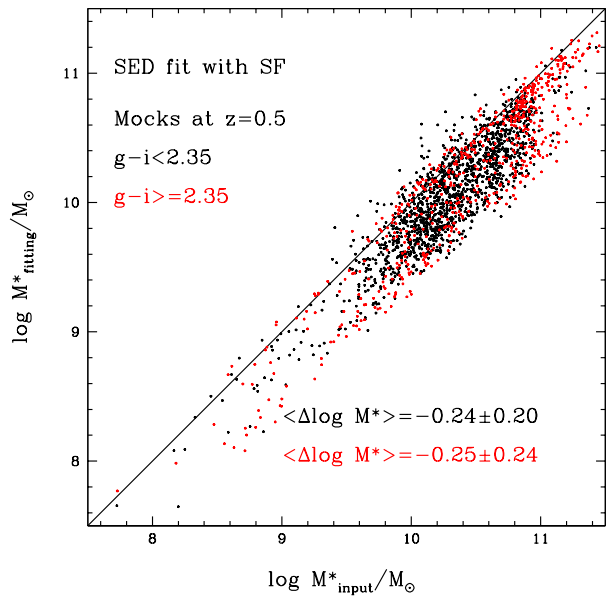}
  \caption{The recovery of stellar mass via SED fitting with the SF template. The mass obtained via SED fitting ($y$-axis) of mock galaxies from semi-analytic models (Tonini et al. 2009) at redshift 0.5 is compared to their intrinsic mass ($x$-axis). The mass discrepancy for blue galaxies ($g-i\lapprox 2.35$, black points) is roughly 0.25 dex.}
  \label{fig:mockSF}
\end{figure}

\begin{figure}
  \includegraphics[width=0.49\textwidth]{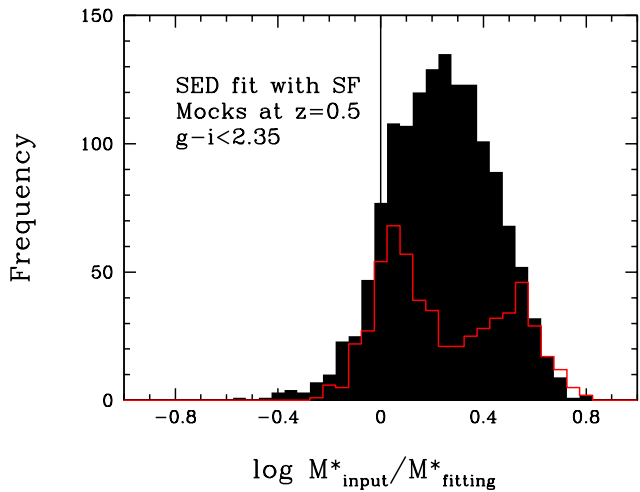}
  \caption{Histogram of mass difference between intrinsic mass and recovered mass, for the SF template applied to blue galaxies, as in Figure~\ref{fig:mockSF}.}
  \label{fig:histodmSF}
\end{figure}

The same experiment for SF galaxies is shown in Figure~\ref{fig:mockSF}. Here we should look at the discrepancy between intrinsic and recovered mass for SF objects (black points). An average offset of 0.25 dex is evident with the intrinsic stellar mass being underestimated by this template (further visualised in Figure~\ref{fig:histodmSF}) For the mass function analysis discussed in the following, we will correct the masses obtained with the SF template by 0.25 dex upward.\footnote{Note that the stellar mass values provided via DR9 have not been corrected.} In Appendix C we discuss the impact of non applying such a correction, on the final mass function.

\begin{figure}
  \includegraphics[width=0.49\textwidth]{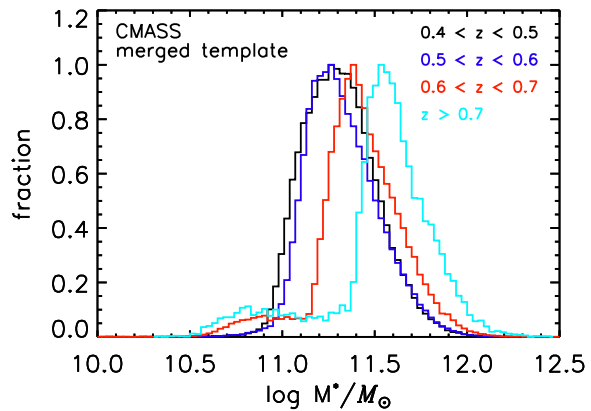}
  \\
  \\
  \caption{The merged template mass distribution for four redshift slices, normalised to the peak mass value in each bin. Here the values of $M^{*}$ obtained with the SF template have been augmented by 0.24 dex as from Figure~14.}
  \label{fig:masszmerged}
\end{figure}
Complementary to Figure~\ref{fig:masszdr9lrg}, Figure~\ref{fig:masszmerged} shows the merged template mass distribution for various redshift slices. The same conclusions hold.

In summary, our mass distribution may still not be the perfect representation of the true stellar masses, but it is anchored to real data through the colour cut and is supported by simulations. Moreover, in a companion paper (Beifiori et al 2013, {\it submitted}) we compare $M^{*}$ with dynamical masses $M_{\rm dyn}$. The two quantities correlate well and $M^{*}$ is never larger than $M_{\rm dyn}$, thereby providing further support to the robustness of $M^{*}$. 

Finally, we have not used a mix of templates for the LOWZ sample, but, as noted in the previous section, this sample has a narrower colour selection hence is populated by a more uniform galaxy population (in terms of star formation properties) with respect to CMASS. Hence the adoption of the LRG template is probably the most appropriate one for the LOWZ sample.

\subsection{Sanity check via emission line statistics.}
As a complementary check, we examined the status of passive or star-forming for BOSS galaxies based on the detection of emission lines. We used the spectroscopic analysis and emission line statistics published in Thomas et al. (2013), (see e.g. their figure 7). For galaxies above the $g-i$~colour cut ($g-i\gapprox2.35$), including the whole CMASS sample, the fraction of emission line is 0.45\%, raising to 4.5\% below this cut. These very low fractions reinforce our proposed  selection.

\section{Comparison to galaxy evolution models}
\label{modelcomp}
\subsection{The semi-analytic model}
\label{mf}
We compare our results with a theoretical light-cone based on the latest version of the Munich semi-analytic galaxy formation and evolution model \citep{guoetal11,henetal12}. These are built on top of the Millennium dark matter simulation that traces the evolution of dark matter haloes in a comoving cubic box $500h^{-1}\rm{Mpc}$ on a side. Merger trees are complete for sub-halos above a mass resolution limit of $1.7\times10^{10}h^{-1}\rm{M}_{\sun}$. A $\Lambda$-CDM WMAP1-based cosmology is adopted (Spergel et al. 2003) with parameters $H_{0}=73~{\rm km \cdot s^{-1}Mpc^{-1}}, \Omega_m=0.25, \Omega_\Lambda=0.75, n=1$ and $\sigma_8=0.9$. 

Baryonic matter forming galaxies is treated as follows. Initial hot gas masses are derived from the mass of corresponding dark matter haloes after collapse, assuming a cosmic abundance of baryons $f_{b}=0.17$. The fate of the gas is then followed through different phases using analytical prescriptions, in particular during cooling and star formation, which may be empirically derived. Feedback from Supernovae II and/or AGNs act to inhibit cooling and - in case of Supernovae - may also reheat the gas, or eject it into an external reservoir. The full evolution history of galaxies - including merging, satellite infall and star formation - is then followed to $z=0$. 
The version of the models used by Henriques et al. (2012) includes AGN feedback as in Croton et al. (2006), the dust model introduced by \citet{delbla07} and the redshift-evolving cold gas-to-dust ratio from \citet{kitwhi07}. This simulation also includes more efficient supernova II feedback and a more realistic treatment of satellite galaxy evolution and of mergers as introduced by Guo et al. (2011).

The spectrophotometric properties of semi-analytic galaxies are obtained using stellar population models. Single-burst or Simple Stellar Population (SSPs) models are assigned to each stellar generation, which is weighted by the mass contribution of the individual star formation episode to the total galaxy mass.  
Henriques et al. (2011; 2012) have updated the De Lucia et al. (2006) and the latest Guo et al. (2011) semi-analytic models with the Maraston (2005) stellar population models, such that now each semi-analytic model is available with multiple choices of input stellar population models. As it has been discussed in the recent literature \citep{tonetal09,fonmon10,henetal11}, the specifics of the stellar population models adopted in the galaxy formation model shape the spectra of model galaxies, which has an important effect on the comparison between models and data. 

The method used to construct the mock catalog is described in detail in Henriques et al. (2012).\footnote{Light cones and data products are publicly available at http://www.mpa-garching.mpg.de/millennium.}
In addition to the pencil-beam format that was originally available, the model is now provided with an all sky light-cone (4$\pi$)  that we will use in this work. The model catalogue is limited to an observed-frame AB \citep{okegun83} magnitude of $i\lapprox 21.0$, significantly deeper than the BOSS limit of $i\lapprox 19.9$. It was constructed by replicating the Millennium simulation box ($500~\rm{Mpc} \cdot h^{-1}$~on a side) with no additional transformations applied. 

The original volume of the Millennium simulation is large enough to sample the most massive galaxies in the Universe, which makes the comparison with BOSS data interesting. Note that the models are normalised to the local mass function, which impacts on the mass of the most massive galaxies that can be found in the simulations.

To make a direct data model comparison we apply to the semi-analytic models the same magnitude colour selection cut that was applied to define the observed sample (the CMASS cut). Here the stellar population model has an effect. The adoption of the Maraston (2005) models instead of the \citet{brucha03} models allow more semi-analytic galaxies to enter the BOSS cut. In the following analysis we shall mostly use the semi-analytic models based on the Maraston (2005) models. 

\begin{figure*}
 \includegraphics[width=\textwidth]{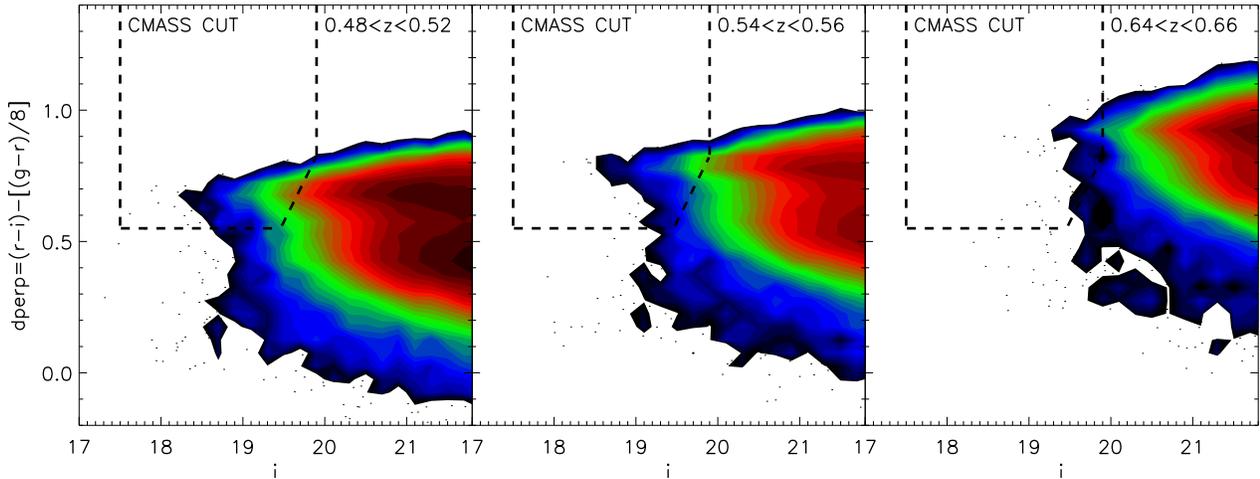}
  \caption{Semi-analytic model galaxies from the model of Henriques et al. (2012) using the Maraston (2005) stellar population models, in the observer-frame $dperp(=(r-i)-((g-r)/8)$ colour vs $i$-mag in the redshift range $\sim~0.5$ to $\sim~0.7$. The CMASS selection cut is shown as dashed lines.}
  \label{fig:cmasscut}
\end{figure*}
Figure~\ref{fig:cmasscut} shows, in the BOSS target selection plot of the observer-frame $i$-mag vs the $dperp$ colour\footnote{dperp is a colour index obtained through the combination of $r,g,i$, such as $dperp=(r-i)-((g-r)/8)$, see Eisenstein et al. (2011).}, the portion of model galaxies entering the CMASS selection cut. Only a tiny fraction of the Millennium galaxies satisfies this selection criterion, because the CMASS cut is designed to select the most luminous and massive galaxies in the Universe (Eisenstein et al. 2011). 

\begin{figure*}
  \includegraphics[width=\textwidth]{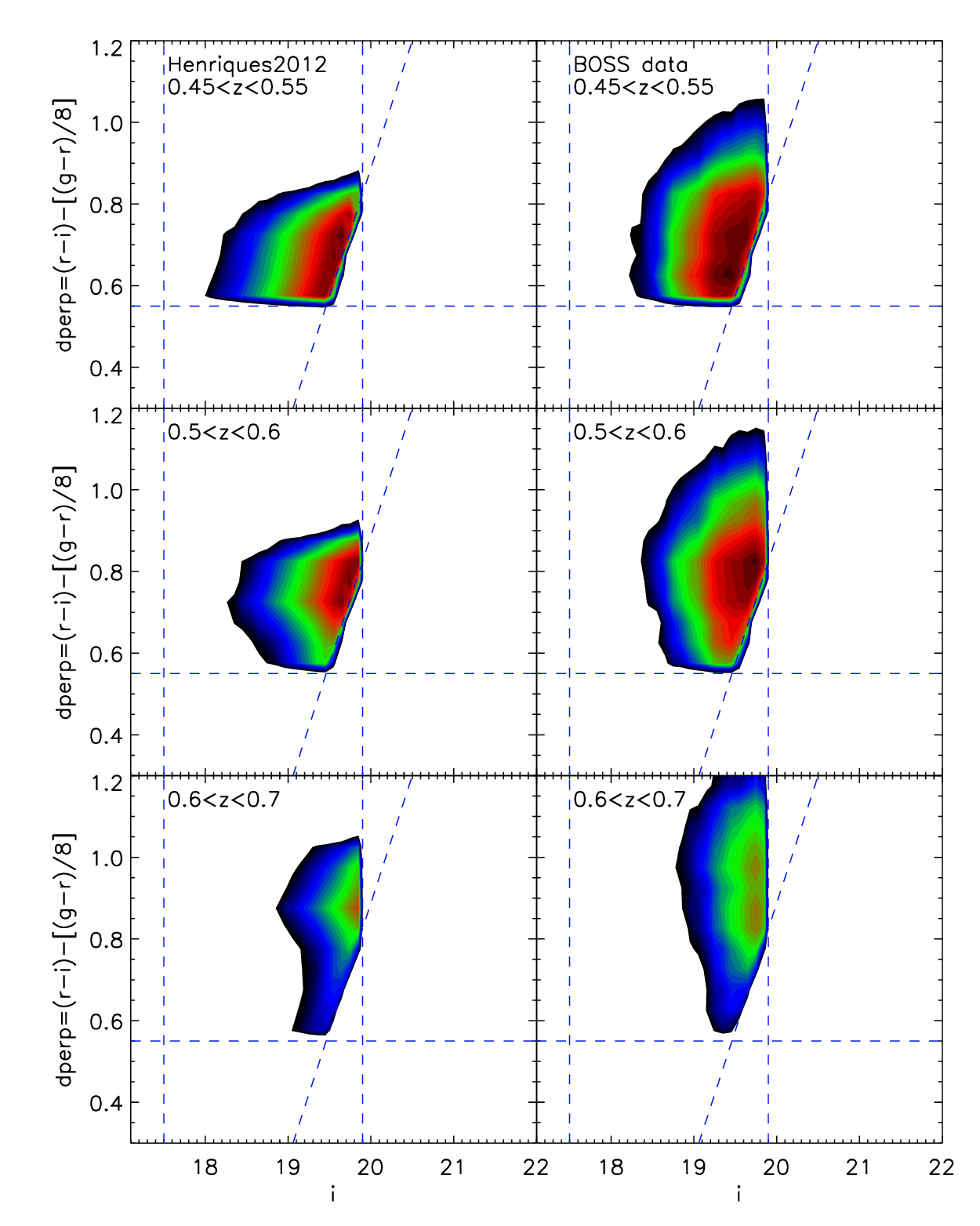}
  \caption{Observed-frame colours of semi-analytic models as in Figure~\ref{fig:cmasscut} ({\it left-hand columns}) and BOSS data ({\it right-hand columns}), in the BOSS selection cut plane of {\it dperp} colour and $i$-magnitude }
  \label{fig:dperpmodeldata}
\end{figure*}

An illustrative approach is to compare the colour distributions of models and data within the target selection cut. Figure~\ref{fig:dperpmodeldata} expands the BOSS selection region in Figure~\ref{fig:cmasscut}. Colours of models and data agree generally well, though one notes a deficit of red galaxies in the models over the entire redshift range. In Section 7.3 we shall discuss this issue in more detail.

\subsection{Stellar mass densities}
\label{sec:mf}
\begin{figure*}
  \includegraphics[width=\textwidth]{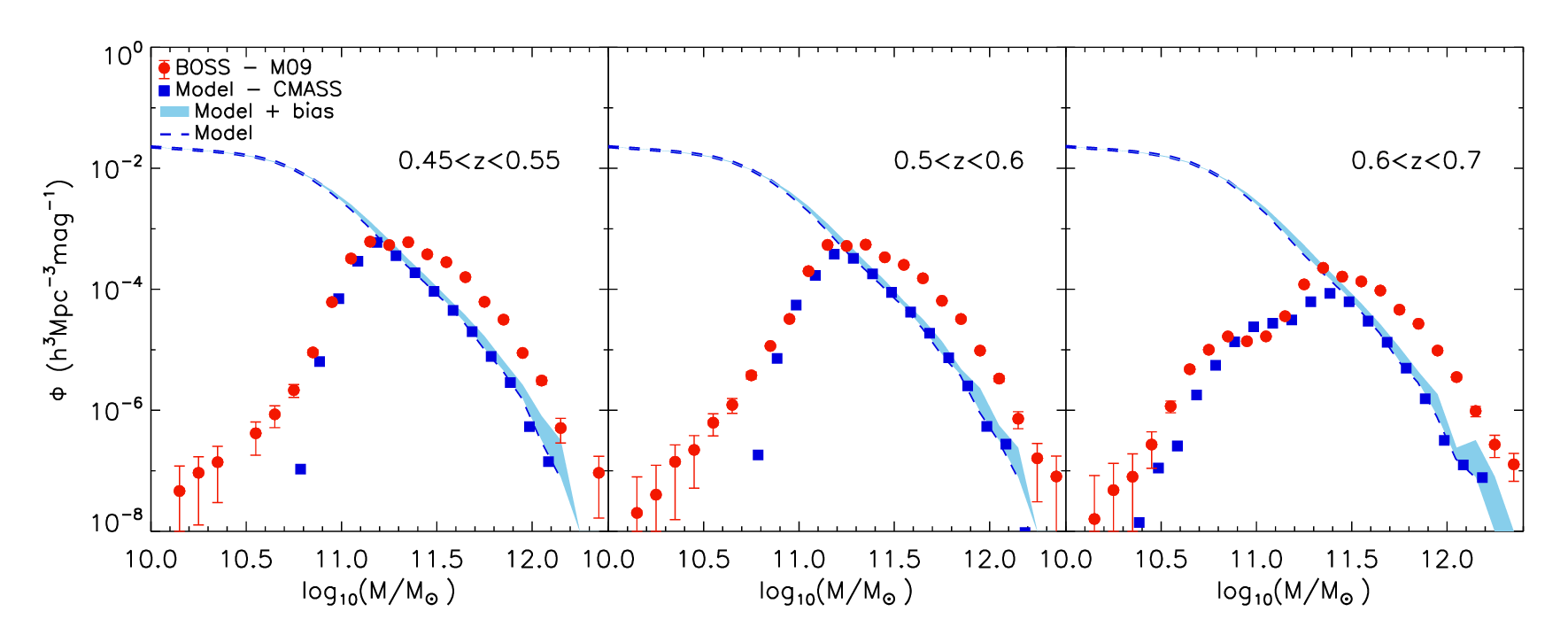}
  \caption{The empirically-derived stellar mass function of BOSS-CMASS galaxies (red points), obtained from stellar masses calculated with the merged template as in Fig.~\ref{fig:massmerged}, for three redshift bins, $0.5, 0.55, 0.65$. Predictions from semi-analytic models (from Henriques et al. 2012 as in previous figures) extracted from a light-cone reproducing the BOSS volume are shown as blue dashed lines. The blue points are the same predictions after application of the CMASS selection cut and an identical mass binning as the data. The light-blue shaded area is the theoretical mass function including a 0.1 dex Gaussian uncertainty in stellar mass derivation from data (bias).}
  \label{fig:massfunc}
\end{figure*}

Figure~\ref{fig:massfunc} displays the stellar mass function of CMASS galaxies (red points with errors), in three redshift ranges. 

These mass functions are calculated based on an effective area (area $\times$ completeness) for the DR9 of 3275 deg$^2$ (see Anderson et al. 2012 for details) and the full volume between the redshift limits i.e. without any further correction applied. We choose such a strategy as our goal is to compare to the semi-analytic model for which we calculate the mass function in the same manner, and it removes any assumptions that would be necessary to calculate the required corrections for the CMASS mass function. Our choice of effective area is driven by the wish to use the exact CMASS catalogue adopted for clustering analysis in BOSS (Anderson et al. 2012). The most important reasons for such a choice is that this sample has been cut to be uniformly selected over  
 the entire survey, so removing any issues of the changing selection  
 over time. It also removes regions of low completeness and is based on  
 the full survey mask including bright star masking etc. (see Anderson et al. 2012 for details). The use of this sample gives us a total number of galaxies of  283819.
 
Error-bars on data-derived stellar masses reflect the $\pm~1\sigma$ variation in stellar mass according to the $\chi^2$~of the fit. 
The errors on the empirical stellar mass function were estimated by combining in quadrature the contributions given by shot noise and by the errors on data-derived stellar masses. The former term was included by using the \citet{geh86} formulation, which takes into account the low-count regime, characteristic of the massive end of the galaxy stellar mass function. The second term is calculated via Monte Carlo simulations, by perturbing individual masses within their errors and recalculating at each iteration the values of spatial density. In particular, by means of this method we obtain spatial-density value distributions for each stellar mass bin, which are used to determine 68\% confidence intervals for each spatial density value plotted in Figures~\ref{fig:massfunc} and \ref{fig:massfuncobs}. The error contribution due to data-derived stellar masses is generally the dominant one, as expected given the large number of galaxies used to measure the stellar mass function, although errors become comparable at the tails of the mass distribution, due to the lower number of objects.

First of all, one should mote the extremely fine resolution in stellar mass at the high mass end and the small error-bars that the BOSS data allow us to achieve. 

The blue lines display the theoretical mass function from semi-analytic models as derived from the full-sky simulations and averaged to the BOSS volume. The blue points are the same simulations where the magnitude-colour CMASS cut has been applied and an identical mass binning as in the data is used. 

The blue shaded area represents a model variance as obtained by accounting for the possible scatter in modelled observations (Baugh 2006, also applied in Fontanot et al. 2009 and Kitzbichler \& White 2009)\nocite{bau06,fonetal09}. This scatter is caused by the fact that several assumptions need to be taken in an empirical mass derivation, such as e.g., the initial mass function, the stellar population model, the wavelength range adopted in the fitting, and the analytical form for the star formation history, including the effects of metallicity and dust reddening. As discussed in previous literature, the consideration of this effect mostly alters the tail of the distribution. 

Usually, a scatter of 0.25 dex in $\log \rm M$~is assumed, as representative of the typical scatter at high-redshift ($z\sim2$, Kitzbichler \& White 2009). 
Here, using our simulations, we can exploit a more quantitative determination for the intrinsic uncertainty in stellar mass. The general template mismatch plus assumed wavelength range\footnote{As in Kitzbichler \& White (2009), we neglect the initial mass function effect, as we use the same IMF in both models and modelled data.} can be read from Figure~\ref{fig:mockLRG}. At the high-mass end, this effect amounts to an asymmetric offset of 0.06 dex, in the sense that our data-derived stellar masses could still be slightly underestimated. On the other hand, there is a scatter of around 0.08 dex, so we decided to translate this result into a Gaussian error distribution of size 0.1 dex to apply to the theoretical mass function. 

The comparison between the red points (modelled BOSS data) and the blue area/blue points (scattered model) in Figure~\ref{fig:massfunc} is then the most appropriate one.
Note that the bias due to the derivation of stellar masses from data could have also been accounted for in the data-derived quantities rather than in the models. We performed this exercise when bootstrapping the observed stellar mass function (as explained earlier in this section). This exercise showed that BOSS observed spatial densities should be corrected towards lower values because of the presence of this bias, which was found to be significant, around 0.1 dex, above $11.8~\log \rm M^{*}/M_{\odot}$, and negligible at lower masses. This effect is equivalent to shifting the theoretical mass function towards higher spatial density values (blue shaded area), in order to reproduce the bias-uncorrected observed mass function. We decided to account for the bias in the models because other data-derived mass functions we shall compare with in Section 5.2.1 (see Fig. 19) do not take this bias into account.

First of all, it is interesting that the models coincide at the massive end independently of whether or not the CMASS cut is applied (compare blue points to blue dashed lines). This result implies that a selection like the CMASS one is perfectly suited to select the most massive galaxies at least from the simulation point of view. In other words, there are no massive galaxies in the models that the CMASS selection would miss.

In Figure~\ref{fig:massfunc} one sees that neither the models nor the data evolve significantly over the BOSS redshift range. This is perhaps not surprising since the redshift spanned is narrow.

Overall, models and data agree fairly well.  There is however a deficit of the most massive galaxies in the models in the mass range $\log \rm M^{*}/M_{\odot} \gapprox 11.6$, of about 0.2 dex, which is uniform over the explored redshift range. This problem was already highlighted in the literature (see next section), but the size of the BOSS sample nails down the result. The turnover in the mass function occurs at slightly different masses, which could result from the different colours of model galaxies and data (see next Section), the photometric errors, or both.

It should be noted that the result of this comparison depends on the details of the stellar mass calculation. For example, the correction of 0.25 dex upward in the value of stellar mass assigned to star-forming objects discussed in Section~4.1 (Figure~15) is the key to get the good match at $10.5 \lapprox \log \rm M^{*}/M_{\odot} \lapprox 11.3$. In Appendix C we show the effect of different assumptions on age cut and templates, on the match between models and modelled data.

The model comparison we present here reaches the highest possible galaxy masses, and cosmic variance, thanks to huge BOSS volume/area, is negligible. We comment on other comparisons of this kind that were previously performed in the literature in the Discussion. We should note that, for the comparison with semi-analytic models, the set of masses for BOSS galaxies we use, whether from this work or from Chen et al. (2012)  does not alter the essence of the conclusions. However, the lower $M^*$~values for BOSS galaxies obtained in this paper (see Appendix A) make the comparison with the models more favourable.

The BOSS data show little evolution within the explored redshift and mass range, but this statement should be taken with caution as we are not dealing with a complete sample; the incompleteness of BOSS is presently not known. For example, note the lower mass density at $\log \rm M^{*}/M_{\odot}\sim11.5$~at the highest redshift bin (right-hand panel) with respect to $z=0.55$, which is the representative redshift for BOSS; this suggests that CMASS is not complete above $z\sim~0.6$ around this mass value, as already argued in Section~4. This results is in qualitative agreement with ongoing simulations of the BOSS completeness (M. Swanson et al. 2013, {\it in preparation}). As we shall see in the next section when comparing with previous results from the literature, the BOSS sample may be not severely incomplete at the high-mass end ($\log \rm M^{*}/M_{\odot}>11.5$) over the entire BOSS redshift range. 

\subsubsection{Comparison with published mass functions}
\label{sec:mf}

The lack of evolution displayed by the field massive galaxy mass function from the BOSS data is in qualitative agreement with earlier results in the literature \citep[e.g.][]{droetal04,bunetal06,cimdadren06,ilbetal10,pozetal10}, including studies considering the luminosity function instead of the mass function in the same redshift range explored here \citep[e.g. ][]{blaetal03,waketal06,cooetal08,lovetal12}.

Our approach, which considers identical volumes in the models and data, should be free from issues related to the unknown completeness of the BOSS sample, and allows us to make a meaningful model-data comparison. Even if the completeness is as yet unknown, it is also instructive to compare our results with the literature in order to estimate where the new BOSS data stand. 
 
\begin{figure*}
 \includegraphics[width=\textwidth]{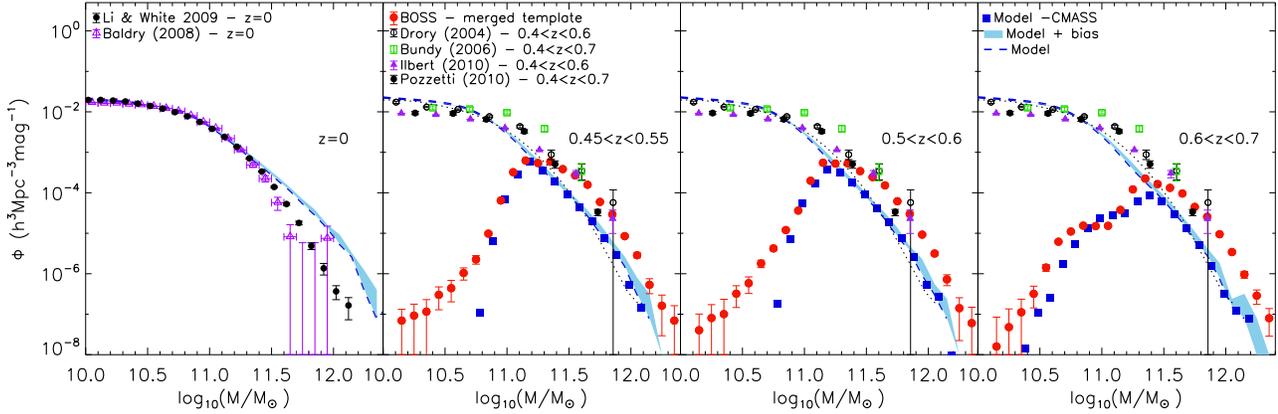} 
  \caption{Similar to Figure \ref{fig:massfunc}, but showing four mass functions from the literature: Bundy et al. (2006, green squares) derived from DEEP2 data; Ilbert et al. (2010, purple triangles) for the COSMOS sample based on photometric redshifts; Pozzetti et al. (2010, black circles), for the $z$COSMOS sample with spectroscopic redshifts; Drory et al. (2004, open circles) from the $K$-band selected MUNICS survey with photometric redshifts. The left panel shows two local $z\lapprox0.1$~mass functions from Li \& White (2009) and Baldry et al.~(2008) as derived from SDSS data.}
  \label{fig:massfuncobs}
\end{figure*}

Figure~\ref{fig:massfuncobs} is identical to Figure~\ref{fig:massfunc}, but with the addition of empirical mass functions derived from other data samples, namely: Drory et al. (2004, open circles), derived from the MUNICS $K$-selected survey with photometric redshift; Bundy et al. (2006, green open symbols) derived from DEEP2 data; Ilbert et al. (2010, purple triangles) for the COSMOS sample using photometric redshifts, and Pozzetti et al. (2010, black filled circles), for the zCOSMOS sample with spectroscopic redshifts. 
There are several other mass functions in the literature, e.g. Borch et al. (2006), Fontana et al. (2006), Bell et al. (2003), but we do not discuss these results as we focus on the high-mass end and explore a high-resolution in redshift binning. In this comparison we need to use works based on a similar stellar initial mass function (IMF) as the one (Kroupa) assumed here. The Bundy et al., Ilbert et al. and Pozzetti et al. mass functions are all based on a Chabrier IMF and Bruzual \& Charlot (2003) stellar population models, while the Drory et al. study is based on Maraston (1998) models and assumed a Salpeter IMF. For plotting the Drory et al. results, we shifted the mass function by $-0.24$ dex, which corresponds to a reduction in stellar mass of a factor 1.6, mimicking the assumption of a Kroupa or a Chabrier IMF.

Also plotted in the left-hand panel of Figure~\ref{fig:massfuncobs} is the $z\sim0$~model mass function along with two local mass functions derived from SDSS-I,II data by Baldry et al. (2008, filled black circles) and Li \& White (2009, open purple triangles). Assumptions on the stellar population model and IMF are the same as in the high-redshift sector. We shall comment on the $z\sim0$~trend in section 5.2.2.

An excellent agreement is found with all previous mass functions. This is remarkable, considering the diversity of data sample, and of methods used to derive stellar masses, both in terms of template models and fitting techniques. The literature works considered here use Bruzual \& Charlot (2003) templates (with the exception of Drory et al. 2004, who adopt Maraston 1998 models) and various wavelength range for the data fitting. As we are dealing with galaxies that are mostly passive and have stellar ages above the AGB period in the Maraston models ($\sim 1$~Gyr), the difference induced by the different template is small (e.g. Maraston 2005; Pforr et al. 2012). The same conclusion was taken in Pozzetti et al. (2010), which tested their results using also Maraston (2005) templates. 

The agreement with Pozzetti et al. also suggests that the use of $u,g,r,i,z$~suffices to obtain robust results with our choice of templates in case of mostly passive galaxies (Pforr et al. 2012), as Pozzetti et al. use a very broad wavelength range extending to the rest-frame near-IR. We plan to test the effect of  near-IR data on our results in a future work (Higgs et al. 2013, {\it in preparation}).

In summary, the BOSS mass function, which extends to $\sim 10^{12} M_{\odot}$, represents the highest-mass mass function published so far in this redshift range in such detail. The comparison with the literature suggests that BOSS may be a complete sample at mass $\gapprox 2\cdot 10^{11}~\rm M_{\odot}$ at redshift below 0.6 and $\gapprox 4\cdot 10^{11}~\rm M_{\odot}$ at redshift above 0.6, which will be verified in future work. 

\subsubsection{Evolution with redshift}
\label{sec:mf}
The mismatch between data and models at the massive end appears to worsen proceeding towards lower redshifts.
The comparisons with $z\lapprox0.1$ mass functions as derived from SDSS data (open black and blue circles) by Baldry et al. (2008) and Li \& White (2008) show that the model overestimates by a larger amount the fraction of massive galaxies. Baldry et al. (2012) confirm - using the GAMA survey - the results they previously obtained using the SDSS. The recent mass function calculated by \citet{mouetal13} using the PRIMUS survey also compares well with the two we plot here.
This evolutionary trend can already be appreciated in Figure~\ref{fig:massfunc}, where one notices that the distance between models and data decreases proceedings towards lower redshifts, and that the amount of massive galaxies at the massive end tends to become slightly larger than in the BOSS data. Worth noticing is also that, instead, for the BOSS sample the agreement between data and models improves from the highest to the lowest redshift range, in particular for galaxies around $\log M^{*}/M_{\odot}\sim 11.5$.

From the model point of view, this result is explained with the secular mass build-up in the hierarchical clustering model. Hence, the model seems to overestimate the evolution with redshift, as also concluded in \citet{almetal08}. Possible solutions to this problem will be mentioned in the Discussion.

Worth noticing is that the density of massive galaxies at redshift 0.5 in BOSS and in the other mass functions plotted in Figure~\ref{fig:massfuncobs} is not consistent with the one for redshift zero derived by the named authors.
 
 This appears to suggest an unphysical {\it negative} evolution with cosmic time, where the density of massive galaxies at high-redshift is higher than at redshift zero. On the other hand, uncertainties in the mass function at redshift zero should also be taken into account.
Li \& White (2009) find a 0.1 dex offset between stellar masses of SDSS galaxies as derived by Kauffman et al. (2003) and Blanton et al. (2007). Chen et al. (2012) re-derive the stellar masses of DR7 galaxies and notice that the new ones are higher (by 0.08 dex) than previously published. Baldry et al. (2008) also discuss the variance between different estimations of the mass function using SDSS data. Interestingly, \citet{beretal10} find a higher mass function at the massive end compared to Li \& White and Baldry et al., as due to a better modelling of the light profile  at the high-mass end, also discussed in Bernardi et al. (2013), and to their choice of model templates to derive stellar masses. In particular, the Bernardi et al. mass function at $z\sim0$~is a factor five higher at the massive end hence in better agreement with hierarchical models. The logical step forward will be to derive the low-$z$ mass function with the same assumptions  for mass calculation taken for BOSS in this paper.

Note that the low-{\it z} empirical mass function is relevant to the models because it is used to normalise the models themselves (Li \& White 2009). The $z\sim0.5$ BOSS data can now be used to calibrate the models over a wider redshift range. 

\subsection{Colours vs mass and the metallicity of galaxies}
\label{sec:colors}
Comparing the spectral energy distribution with the stellar mass, is a powerful approach to gain insight into the galaxy evolution process, as the SED records the history of star formation, e.g. the age distribution and the metallicity, which encodes information about merging and gas infall histories and feedback processes. Here we use the SDSS colours which at the BOSS redshift mostly sample the rest-frame optical, although towards the lowest boundary in redshift the $i,z$ bands record a touch of the rest-frame near-IR.

\begin{figure*}
  \includegraphics[width=\textwidth]{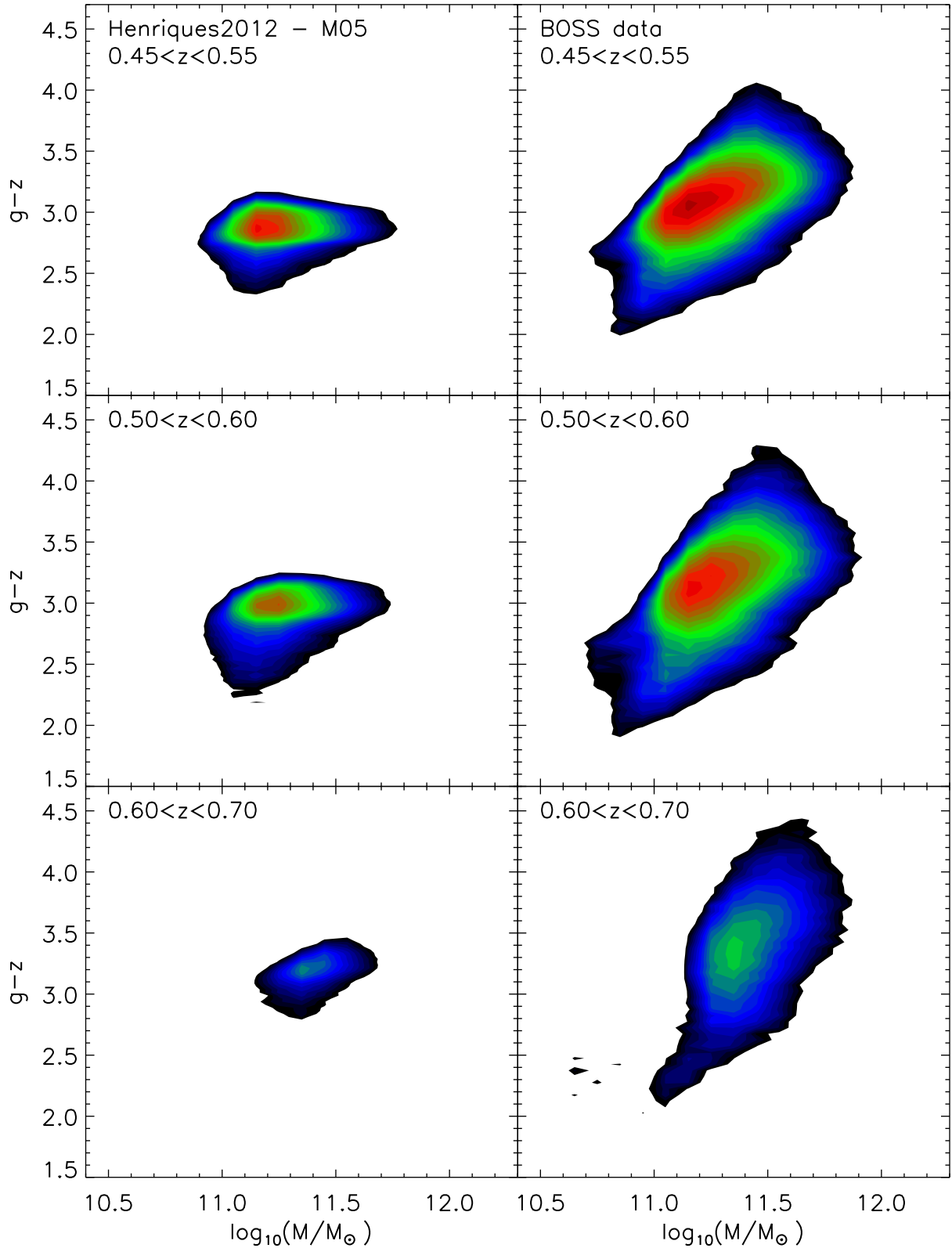}
  \caption{$g-z$~observer-frame colour vs stellar mass for BOSS/CMASS galaxies in the redshift range 0.45 to 0.7 (right-hand column). Equivalent relations  from semi-analytic models are shown in the left-hand column. The data display colour-mass relations with the most massive galaxies being the reddest, which are not seen in the models. Diagrams for other SDSS colours are shown in the Appendix.}
  \label{fig:colmass}
\end{figure*}

Figure~\ref{fig:colmass} shows the relations between the observed-frame colour $g-z$ and the stellar mass, for BOSS CMASS galaxies and semi-analytic models (right-hand and left-hand panels, respectively), in three redshift slices. The number counts under each contour have been weighted by the volume of each catalog. The models by Guo et al. (2011) - modified by Henriques et al. (2012) as to include the M05 stellar population models - are used, as in previous Sections. Similar plots using other colours are listed in the Appendix. Here we discuss this specific colour as it samples the same rest-wavelength of $u-i$, which was used in Guo et al. (2011), and with which we shall compare later in this Section. 

Focussing on the data first, we see that the BOSS galaxies display the well-known colour-magnitude - here colour vs mass - relation, where larger galaxies are redder (e.g. Bower, Lucy \& Ellis 1992). This qualitatively holds for all examined colours (see Appendix). In the local Universe, the colour-magnitude relation is interpreted in terms of metallicity, with the most massive galaxies being the most metal-enriched \citep{kodari97}. This is confirmed by the detailed analysis of the metal content of galaxies through absorption-line modelling (e.g. Thomas et al. 2005; 2010). Moreover, \citet{kodetal98,kodetal99} and \citet{staeisdic98} show that colour-magnitude relations similar to those in the local Universe exist for galaxies in clusters at redshifts comparable to BOSS up to $z\sim 0.9$. \citet{cooetal06} analyse a sample of 20,000 massive SDSS galaxies up to redshift 0.3 and show that such relations exist for field galaxies, dependent of the band, although those for field galaxies show a $10\%$~larger dispersion than those in clusters. 

Here we demonstrate that well-defined colour-mass relations hold for field galaxies at the BOSS redshifts. Since the BOSS sample is dominated by high-mass galaxies, which, in terms of stellar content and chemical enrichment, do not differ much from their counterparts in the field (Thomas et al. 2010, Peng et al. 2010), we do not expect that these relations would be very different from those for cluster galaxies.
We are unable to plot a mass-metallicity relation in our paper as the metallicity derived through broad-band SED fit is not well-resolved, and moreover the LRG model that is used for most galaxies has a fixed metallicity. The analysis of the absorption lines in BOSS galaxies stacked spectra will be developed in a parallel paper (Thomas et al. 2013, {\it in prep.}).

Galaxy colours in the models do not vary as a function of stellar mass, in other words, the colour-mass relation in the models is flat, and the model colours are typically bluer than the real galaxy colours.\footnote{Note that - as we use the observed frame where colours get redder because of redshift - the large span in the observed colours may come from uncertainties in redshift. The size of the mismatch between data and model is probably exacerbated by this effect.}
As is well known, galaxy colours can vary as a function of age, metallicity or dust content.  Dust effects  should play a minor role, as the bulk of the massive CMASS galaxies are not very dusty, as already discussed (see Section 3.2).

A substantially younger age component in the models - which causes colours to remain blue - is also not the main driving of this mismatch as - at redshift 0.5 - the galaxy ages in the present semi-analytic models are strongly peaked at old ages, with a very low percentage scattering to low ages (Henriques et al. 2011, Figure 5). This conclusion would not be the same for other semi-analytic models, as the same Figure shows. 

We are left with metallicity effects as a possible explanation. It is known that galaxies in semi-analytic models are generally quite metal-poor even at high-masses; their metallicity barely reaches the solar value as discussed e.g. by \citet{pipetal09}, \citet[][, their Figure 10]{hentho10} and also briefly pointed out in Tonini et al. (2009) and Pforr et al. (2012). Moreover, \citet{saketal11} describe the difficulty in matching the mass vs gas-phase metallicity relation at high-redshift even when implementing a sophisticated recipe for chemical enrichment. We shall return to this point for the discussion.

We also should comment on the effect of population synthesis models. We checked that the use of the BC03 population models makes only a marginal difference in the semi-analytic model predictions in the SDSS bands, which sample a rest-frame spectral region, between 3400 \AA\ and 6400 \AA, which is not vastly different between the two models, especially because the model galaxies are mostly old and have roughly half-solar metallicity. The choice of population synthesis model appears to matter, however, at higher metallicity, as we discuss below. 

Guo et al. (2011) perform a similar analysis as in Figure~\ref{fig:colmass}, by comparing the rest-frame $u-i$~galaxy colours in bins of stellar mass at redshift zero, using SDSS data.  Models and data are found to compare remarkably well for galaxies with masses in the range $\rm \log M^{*}/M_{\odot}\sim~9.5-10.5$\footnote{At lower masses, the models are redder, which - as discussed by the authors - is due to substantial fraction of dwarf satellites (roughly half) in the models which finish their star formation early and become passive. The observed fraction of such passive dwarfs is substantially smaller.  Our data do not encompass this low-mass range hence we cannot address further this problem.}. At the high-mass end, $\rm \log M^{*}/M_{\odot}\gapprox 10.5$, model galaxies are found to be bluer and to span a narrower colour range with respect to the data. 
The discrepancy discussed by Guo et al. is identical to the one we point out in Figure~\ref{fig:colmass} for galaxies at redshift $\sim 0.5$. 
Galaxy metallicities at redshift zero are centred around 0.5 $Z_{\odot}$. This value is smaller than what is inferred by observational data using stellar population modelling of absorption lines \citep{thoetal05,thoetal10,galetal06,smietal09}, as discussed by Henriques \& Thomas (2010).

Hence, our conclusion is that the main cause of the discrepancy between models and data for the colours of massive galaxies lies in the metallicity, which is too low in the models. Guo et al. (2011) conclude the opposite, namely that metallicity/age effects are unlikely to be able to explain this discrepancy. This conclusion is based  on the evidence that the $u-i$~colour of the Bruzual \& Charlot (2003) models for 12 Gyr and twice solar metallicity (and a Chabrier IMF) is at most 3.07, whereas the peak of the data is around 3 and extends up to $\sim$~3.5. On the other hand, the equivalent model from Maraston 2005 (for a Kroupa IMF) has $u-i=3.47$\footnote{See www.maraston.eu}. Hence, the semi-analytic models with a higher metallicity for the galaxies and using the M05 stellar population models could match the colours, for metallicity values - between solar and twice solar - that are in accord with what is derived observationally. This finding further stresses the importance of evolutionary population synthesis for the theoretical modelling of galaxies (Tonini et al. 2009; Henriques et al. 2011; Monaco \& Fontanot 2010).

The conclusion from this section is that the most massive galaxies in the models need to be more metal-rich to match the observations.

\section{Summary and Discussion}
\label{concl}
We have calculated the photometric stellar masses for galaxies in the BOSS survey from the commissioning stage through the first release of data to the public (DR9). We have used the BOSS spectroscopic redshift and standard SDSS photometry $u,g,r,i,z$, to perform broad-band spectral energy distribution (SED) fitting with {\it HyperZ} (Bolzonella et al. 2000) using various galaxy templates. In particular, we exploit our previously published Luminous Red Galaxy (LRG) best-fitting template (Maraston et al. 2009), which is composed of a major metal-rich population containing traces ($3\%$~by mass) of metal-poor stars, both populations being coeval and in passive evolution. This template provides a good description of the redshift evolution of the $g,r,i$ colours of LRG galaxies in the redshift range 0.3 to 0.6 from the 2SLAQ survey (Maraston et al. 2009; see also Cool et al. 2008 who used a preliminary version of the same template). This template was also used to design the target selection for BOSS (Eisenstein et al. 2011). Furthermore, as the BOSS target selection includes galaxies that are bluer than the classical LRGs, we also use a template suite allowing star formation, ranging from standard $\tau$-models to constant star formation and spanning a wide metallicity range (from 0.2~solar to twice solar). For both templates we employ a Salpeter (1955) as well as a Kroupa (2001) Initial Mass Function (IMF) and consider the mass lost via stellar evolution. 

Independently of the adopted template, the result is that BOSS galaxies are massive and display a narrow mass distribution, which peaks at $\rm \log M/M_{\odot}\sim11.3$~for a Kroupa IMF. We also study the uniformity of the mass sampling as a function of redshift and find that BOSS is a mass-uniform sample over the redshift range 0.2 to 0.6 (see also White et al. 2011). Qualitatively speaking, incompleteness emerges at redshift above 0.6 and $\log \rm M^{*}/M_{\odot} \lapprox 11$.

The galaxy stellar mass depends on the adopted template, and generally it is not obvious which template is the best choice as the galaxy star formation history is not known. To make a robust template choice is especially difficult for large galaxy databases, in which objects cannot be handled on an individual basis. For obtaining a unique set of reference stellar masses, we adopt an empirical colour cut developed in a companion paper (Masters et al. 2011) which is able to separate galaxies with early-type morphologies from later-type ones at redshift above 0.4. We then use the stellar masses obtained with the LRG passive model for galaxies on the 'early-type side'  of the colour criterium, and the values obtained with the star-forming template for galaxies on the 'late-type' side.  In this way we obtain a merged mass distribution in which the assignment of the stellar population template is motivated by the observed galaxy morphology and colours. 

Noticeably, we also study using mock galaxies how well the chosen template is able to recover the true stellar mass. Based on the results, we apply a correction of 
$+0.25$ dex to the stellar mass for the bluest (star forming galaxies) and we use an age cut of 3 Gyr as a limiting fitting age for the reddest and passive galaxies. The effects of these priors on the conclusions on galaxy evolution are shown in detail in an Appendix.

The BOSS galaxy sample used here, comprising $\sim 400,000$~massive galaxies at redshifts $\sim 0.3-0.7$, is ideally suited to study at unprecedented detail the evolution of the most massive galaxies at late epochs. We compare the mass distribution and the colours of BOSS galaxies with predictions from semi-analytic models of galaxy evolution based on the Millennium simulations (Guo et al. 2011; Henriques et al. 2012). The simultaneous comparison of mass and colour is crucial. These quantities in the models are affected by the prescription for AGN feedback \citep{guoetal11,delbla07,croetal06,catetal05}, which is likely far too simplified, and probably incorrect in detail (Bower et al. 2012). 

To perform a robust comparison free as much as possible from possible completeness issues, we consider the models in light-cones using the BOSS effective area and the target selection cuts. The large area of the BOSS survey and the selection cut at the high-mass end allow us to pose results on an unprecedentedly solid statistical ground.

Overall the models perform fairly well in comparison with the data in terms of stellar mass density distribution at redshift $\sim 0.5$. This is already visible in previous work (cfr. Figure~20 by Pozzetti et al. 2010). However, the density of the most massive galaxies, $\log \rm M^{*}/M_{\odot}\gapprox 11.4$, is larger in the data compared to the models over the explored redshift range. This discrepancy increases down to redshift zero, as the models grow progressively bigger galaxies consistently with the hierarchical mass build-up. These conclusions are qualitatively consistent with those taken in previous articles (Fontanot et al. 2009, Pozzetti et al. 2010, Ilbert et al. 2010), who noticed that the evolution at the high-mass end of the empirical mass function is much milder than the one at the low-mass end, in agreement with the {\it baryonic mass downsizing}. On the contrary, the models display an {\it up-sizing} where the massive end and especially the passive population \citep{catetal08,fonetal09} evolves faster with respect to the low-mass end. Due to the BOSS target selection we can only reach conclusions about the high-mass end here, but we are able to extend the analysis to the very massive end that was not probed previously.

The extension to high mass is crucial for understanding the evolution of the most massive galaxies with respect to galaxy formation models. For example, Bower et al. (2006) conclude that the predicted mass function in their semi-analytic models reproduces reasonably well the observations all over the redshift range from zero to five. Examining their Figure 6, however, one notices that their model at redshift 0.5 lacks the most massive galaxies compared to our BOSS results and to the semi-analytic models we use here. Bower et al. could use only observed mass functions that extended up to $\sim 10^{11}~M_{\odot}$. 

Almeida et al. (2008) on the other hand noticed that the observed luminosity function of LRG at $z\sim 0.5$ is not matched by either the Bower et al. (2006) or the Baugh et al. (2005) semi-analytic model of galaxy formation and evolution. The Bower et al. model is successful at predicting such abundance at lower redshift ($z\sim0.24$). This implies a different redshift evolution in the models and the data similar to what we find here. The models we use in this work appear to be more successful at redshift 0.5 than at lower redshift, as already discussed in the literature. 

As star formation is quenched by AGN feedback in these models, the secular evolution of massive galaxies is mostly
determined by mergers, particularly by minor mergers, since for the most massive galaxies the mass ratio to other galaxies is always large. The relative growth of
the mass function between $z$=0.5 and $z$=0 is therefore strongly affected by
the treatment of the physics of satellite galaxies. In particular, tidal disruption of stellar material can significantly
decrease the amount of mass accreted onto massive galaxies, and move it
into the intra-cluster light \citep{monfontaf07,hentho10}. A more effective
implementation of this process could help in reducing the excessive build up
of massive galaxies in the Guo et al. (2011) models and ease the tension with $z$=0 data.

We find that our light-coned mass function compares well with the mass function based on the $z$COSMOS survey(Pozzetti et al. 2010). 

The comparison with these previous analysis suggests that BOSS is a complete sample at mass $\gapprox 2\cdot 10^{11}~\rm M/M_{\odot}$ at redshift below 0.6 and $\gapprox 4\cdot 10^{11} ~\rm M/M_{\odot}$ at redshift above 0.6. These suggestions will be verified quantitatively in future works.

The BOSS mass function at $z\sim0.5$~appears to be in tension with local mass functions in giving a higher number of massive galaxies at high redshift with respect to redshift zero. This tension is also seen in previous works. On the other hand, the most recent re-determinations of the massive end of the local galaxy mass function (Bernardi et al. (2010)) give a higher mass density at the massive end, in better agreement with BOSS and the other high-$z$ works. This is clearly a promising result to follow up.

In summary, the BOSS mass function which extends up to $\sim 10^{12} M_{\odot}$~represents the highest-mass mass function published so far in this redshift range in such detail in redshift and mass. 
BOSS now offers an interesting data base of massive galaxies for calibrating models of galaxy formation and evolution at the highest mass end at high-redshift which is free by cosmic variance and small-number statistics.

A comparison of the colours of BOSS galaxies and models demonstrates that BOSS galaxies define colour-mass relations similar to those of local galaxies, with colours becoming redder with stellar mass. The models, however, span a narrower (bluer) colour range, and in particular their colours do not vary with stellar mass, i.e. the models do not display a colour-mass relation. We argue that the main driver for this discrepancy is the metallicity, which in the models is too low, a conclusion which is consistent with evidence from other work in the literature. Interestingly, Guo et al. (2011) discarded this possibility when comparing - in a similar fashion as we do here - SDSS galaxies with models at redshift 0. Their conclusion is based on the evidence that Bruzual \& Charlot (2003) population synthesis model colours do not  vary enough as a function of metallicity as to encompass the observed colours. On the other hand, the Maraston (2005) model colours show a stronger variation with metallicity (between solar and twice solar) which would just be appropriate to reconcile the models with the data. In summary, an improvement to the models should go in the direction of gaining a higher metallicity for the most massive galaxies.

The low metallicity of massive galaxies may be more a problem of semi-analytic models than galaxy formation in general. In fact, chemical enrichment in hydro-dynamical simulations proceeds more efficiently than in semi-analytic models and galaxies reach higher metallicities (Dave', Finlator and Oppenheimer 2006; Naab et al., {\it in preparation}; Dave' et al. 2012; Cattaneo et al. 2011)\nocite{davfinopp06,davfinopp12,catetal11}.
On the other hand, semi-analytic models are still the most efficient approach for large galaxy simulations, hence the goal should be to improve upon the star formation, chemical enrichment and feedback in semi-analytic models of galaxies. Moreover, it may be the full hierarchical growth, in terms of satellite accretion and gas infall, which is responsible for diluting the metallicity (Henriques \& Thomas 2010), which is not yet included in full hydro-dynamical simulations. Much effort is currently invested in galaxy formation science and the next few years will certainly see major step forward towards the solution of these problems.

\section*{Acknowledgments}
Funding for SDSS-III has been provided by the Alfred P. Sloan Foundation, the Participating Institutions, the National Science Foundation, and the U.S. Department of Energy Office of Science. The SDSS-III web site is http://www.sdss3.org/.
SDSS-III is managed by the Astrophysical Research Consortium for the Participating Institutions of the SDSS-III Collaboration including the University of Arizona, the Brazilian Participation Group, Brookhaven National Laboratory, University of Cambridge, Carnegie Mellon University, University of Florida, the French Participation Group, the German Participation Group, Harvard University, the Instituto de Astrofisica de Canarias, the Michigan State/Notre Dame/JINA Participation Group, Johns Hopkins University, Lawrence Berkeley National Laboratory, Max Planck Institute for Astrophysics, Max Planck Institute for Extraterrestrial Physics, New Mexico State University, New York University, Ohio State University, Pennsylvania State University, University of Portsmouth, Princeton University, the Spanish Participation Group, University of Tokyo, University of Utah, Vanderbilt University, University of Virginia, University of Washington, and Yale University.\\
Numerical computations were performed on the Sciama High Performance Compute (HPC) cluster which is supported by 
the ICG, SEPNet and the University of Portsmouth.\\
CM acknowledges relevant discussions with Ivan Baldry, Andrea Cimatti, Olivier Le F\'evre, Danilo Marchesini and Lucia Pozzetti on the galaxy mass function, and with Matt Auger and Michele Cirasuolo on stellar mass determinations. A special thank to Micol Bolzonella for her help with the use of the HyperZ software. A special thank to Harry Worsfold, the Nuffield Bursary Project Placement student who helped us compiling the bibliography. Finally, we acknowledge a very careful Referee for insightful comments that helped clarifying the paper.
\appendix
\section{Comparison with other stellar mass calculations in DR9.}
\label{masscomp}
Chen et al. (2012) calculate stellar masses for BOSS galaxies using the individual BOSS spectra and a procedure based on Principal Component Analysis (PCA) for obtaining the star formation history of the galaxy from spectral fitting. The PCA is run on a library of stellar population models for a variety of ages, metallicities and dust content to identify its principal components over the rest-frame wavelength range $3700-5500$~\AA. 

Chen et al.  present results based on both the Bruzual \& Charlot (2003) and the Maraston \& Str\"omback (2011) stellar population models\footnote{The Maraston \& Str\"omback (2011) stellar population models are the high-resolution version of the \citet{CM05} we adopt here for the star forming template, and use empirical stellar libraries, as in the LRG model.}. Chen et al. assess the dependence of their results on the different stellar population models. There is a constant offset of $0.12$~dex, mostly concentrated at low galaxy ages, in the sense of lower stellar masses obtained with the Maraston \& Str\"omback (2011) models. This difference is most likely due to the different energetics and temperatures in the phase of Red Super Giant in the stellar evolution models adopted in the two population models (see Chen et al. 2012). This offset is smaller than the $0.2-0.3$~dex usually reported in the literature for stellar masses obtained from SED fitting using Bruzual \& Charlot and Maraston models (e.g. Ilbert et al. 2010). The offset can be due to a combination of the following two effects. First, BOSS galaxies are generally older than the AGB ages ($\sim1$~Gyr) where the two models mostly differ. Second, the wavelength range adopted in the fit does not include rest-frame near-IR wavelengths where the two models differ the most.

\begin{figure}
  \includegraphics[width=0.49\textwidth]{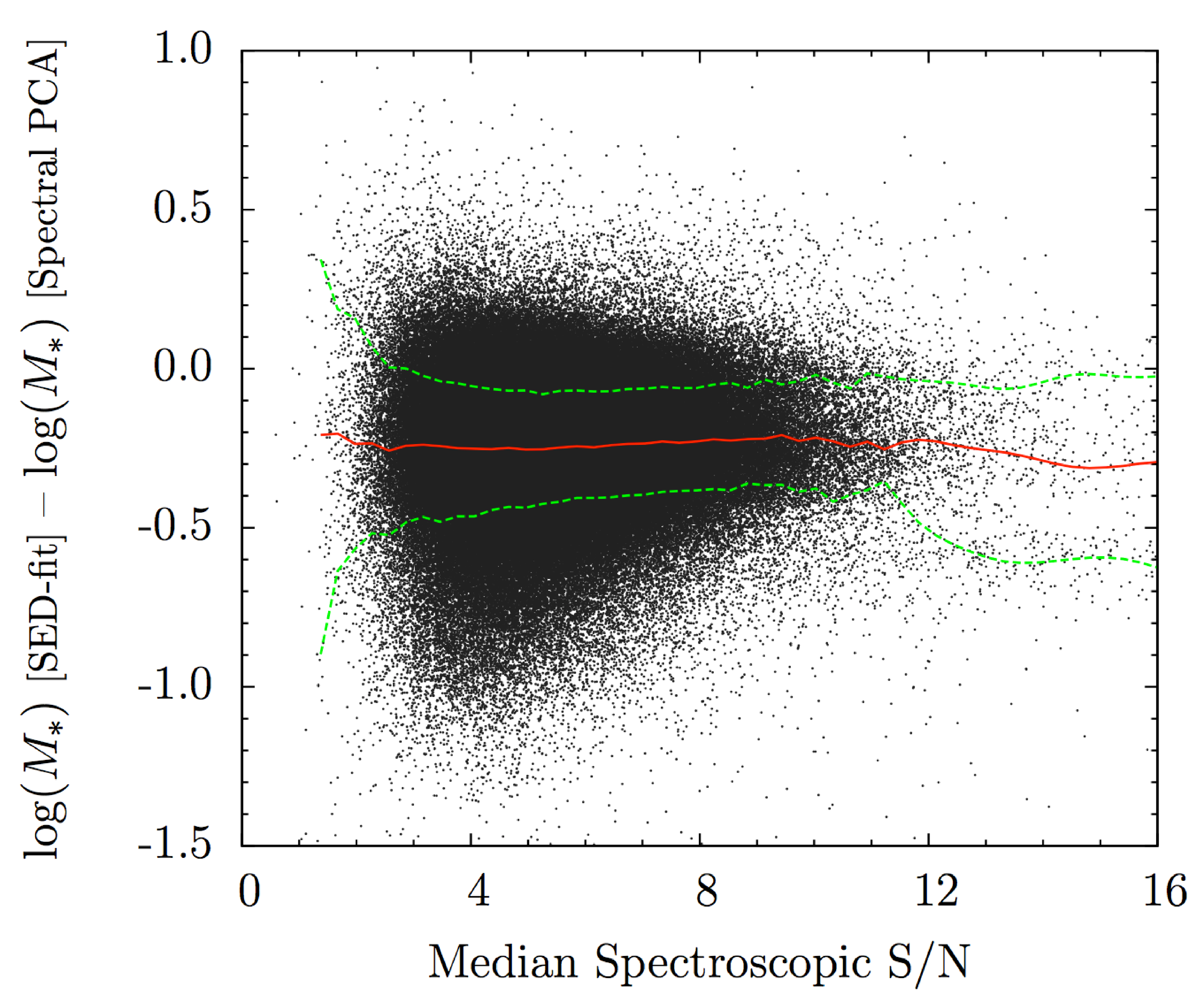}  
  \caption{Difference in $M^{*}$ for CMASS galaxies between masses from this work obtained via broad-band SED fit of $u,g,r,i,z$~and those from Chen et al.~(2012) based on PCA spectral fitting of individual BOSS spectra, as a function of the median spectroscopic S/N in the spectral window $3700.57 - 5498.80$~\AA. The red line highlights the median of the difference and the two green lines the $\pm 1~\sigma$ variation.}
  \label{fig:compmasschen}
\end{figure}
Here we focus on the dependence of stellar mass on the two methods, namely high-resolution spectral fitting versus broad-band SED fit. Hence we focus on the comparison at fixed population model and we adopt Chen stellar masses based on the Maraston \& Str\"omback (2011) models. 
Figure~\ref{fig:compmasschen} shows the difference in stellar mass between the values of $M^*$~ derived in this work and those by Chen et al. (2012), both based on Maraston's models. The difference is shown as a function of the median spectral S/N\footnote{The different absolute scale of S/N in Figure~\ref{fig:compmasschen} compared to Figure 12 of Chen et al. 2012 is due to the fact that here we use the S/N in the spectral window $3700.57 - 5498.80$~\AA, whereas Chen et al. used the S/N determined over the entire spectrum. The trend of the comparison is not affected by this choice.}. A constant offset of 0.2~dex is evident, with the spectral masses being larger than our photometric ones. This difference is independent of the S/N. 

Also Chen et al. (2012) find that spectral stellar masses, at BOSS S/N, are higher, by 0.1~dex, than those they derive from broad-band SED fitting on $g,r,i,z$, using the same model templates. 

Still, the discrepancy we find ($\sim0.2$ dex) is larger than the one quoted by Chen et al. (2012). Here there is another factor entering, namely the model star formation history. We use a mostly passive template and do not include reddening from dust in the fitting, while Chen et al. include star formation and dust. While the mere use of the passive template should push the analysis to higher masses (as the M/L~of stellar population models increases with age), the inclusion of dust may force the model to fit for a larger old component  than in case of a single age template to balance the younger and dusty component. This increases the global M/L ratio, hence produces a higher $M*$\footnote{This is the opposite effect reported by Maraston et al. (2010) and Pforr et al. (2012), who find that when dust is included, $M^{*}$~decreases because dust favours young solutions with a low M/L. However, this result holds for single-age fitting, while Chen et al. consider a composition of populations and in this case exactly the opposite effect happens.}. A similar conclusion is drawn in Chen et al. (2012), who show (their Figure 13) that when dust is excluded, their $M*$~is reduced by $\sim 0.08$~dex. It is suggestive that - using emission line information - Thomas et al. 2012 (Figure 8) find very little dust in the reddest CMASS galaxies.

Hence, the different priors used in constructing the two model libraries and the low S/N of the BOSS data appear to explain the discrepancy in stellar masses.

Nonetheless, we explored two further possible sources of difference that can affect the stellar mass derivation.
First, the PCA-spectral stellar-mass-to-light ratios derived by Chen et al.~(2012) are based on the light which falls within the 2 arc-second SDSS fiber and translated into total galaxy masses by multiplying the derived M/L~ratio by the light (in the $i$-band) derived from {\it cmodelmag}.  As already pointed out by Chen et al., this approach assumes that the M/L~is constant over the whole galaxy. However, if galaxies have colour gradients that are detected by the data, the total M/L will not be the same as the M/L ratio within the fiber. To quantify this effect, we perform SED fit using fiber-magnitudes, after scaling them to the brightness of the {\it i}-band {\it cmodelmag} as in our standard procedure. 
\begin{figure*}
  \includegraphics[width=0.49\textwidth]{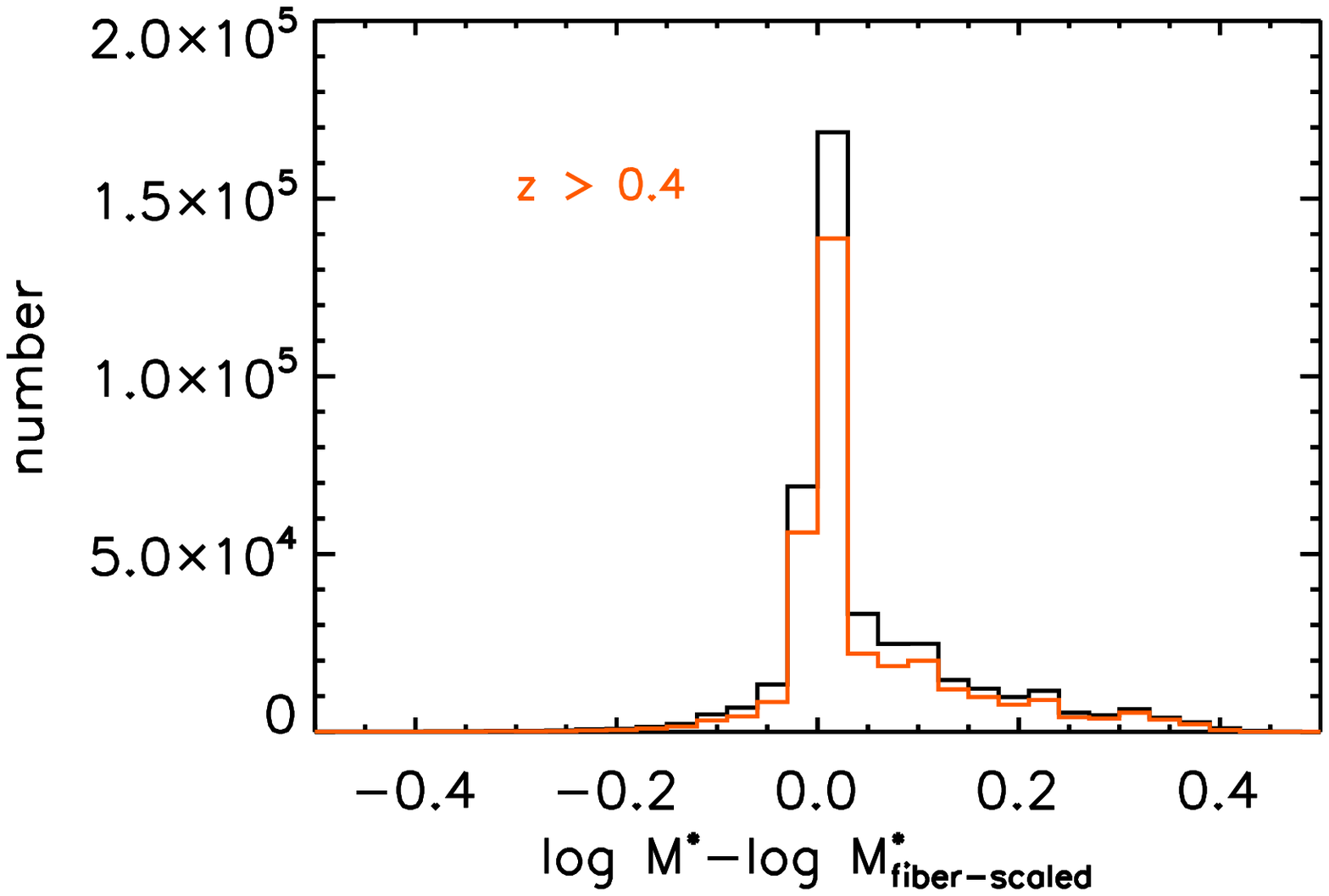}
   \includegraphics[width=0.49\textwidth]{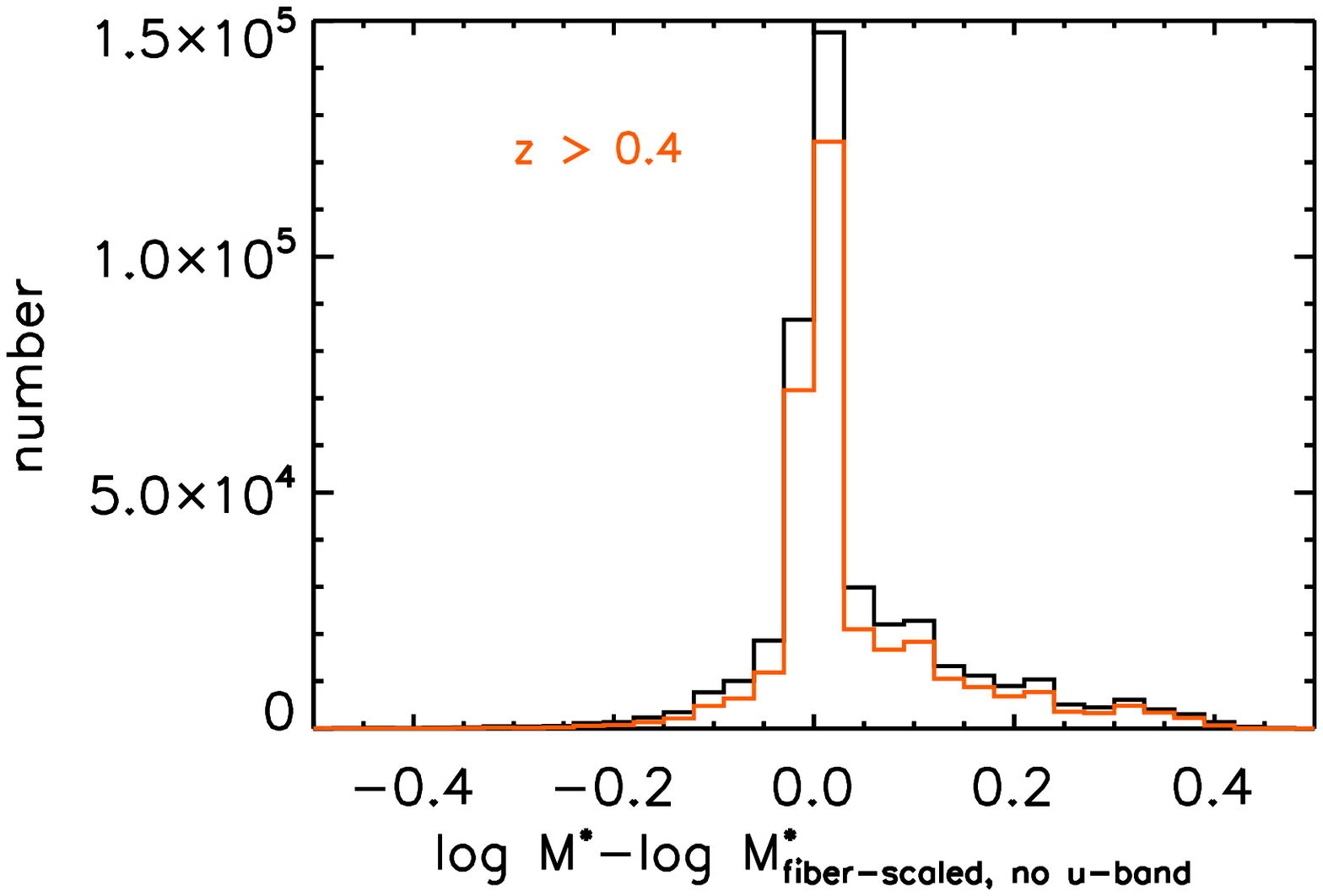} 
  \caption{Difference between the total stellar mass obtained with total magnitudes and the 'fiber' stellar mass scaled to the total light ({\it left-hand panel}) and the same, but additionally excluding the $u$-band from the SED fit for the 'fiber-scaled' stellar masses ({\it right-hand panel}). Galaxies in the high-$z$ CMASS sample are shown in red.}
  \label{fig:compmassfib}
\end{figure*}

Figure~\ref{fig:compmassfib} ({\it left-hand panel}) shows that there is indeed a slight difference between the two mass estimates - true total mass minus the total mass obtained from the fiber magnitudes scaled with the total luminosity. The total masses are slightly larger than the {\it fiber-scaled} ones (mean of 0.044 dex, with dispersion of 0.1 dex, and 0.046 dex with dispersion of 0.098 dex for the high-$z$~sample, red histogram in Figure~\ref{fig:compmassfib}). This is due to slightly larger ages obtained using total magnitudes. Hence, this effect cannot explain the offset with the Chen et al. masses, because their masses are larger than ours. However, this trend refers to our photometric SED-fit, and it may be different in case of spectral fitting. 

The second effect that may be acting to cause the mass discrepancy is related to the fact that we include the $u$-band in the SED fit, while Chen et al. do not. We repeated our calculations by excluding the $u$-band, but the results hardly change (Figure~\ref{fig:compmassfib}, {\it right-hand panel}). The mean of the distribution is 0.038 dex, with standard deviation 0.11 dex (and mean of 0.039 dex with standard deviation of 0.098 for the high-$z$~sample).

In summary, we investigated and discussed the sources of difference between stellar masses from broad-band SED fit and those derived via spectral fitting of individual spectra. From Chen et al. one sees that - due to the limited quality of BOSS data - the mass obtained via spectral fitting is 0.1 dex higher then the SED-fit masses. In addition, the different priors used in constructing the model libraries  push the spectral-based stellar masses towards higher values. The sum of these effects can explain the difference between the spectral masses and our SED-fit masses. 

\section{Model rest-frame magnitudes of BOSS galaxies.}
\label{magskcorr}
The fitting of theoretical templates to derive galaxy stellar masses allows us to obtain other interesting quantities. Using HyperZ, we generated the rest-frame magnitudes in $u,g,r,i,z$~of the best-fitting template for all BOSS galaxies. These are the magnitudes each galaxy has according to the best-fit template in its rest-frame, e.g., $M_{r}$~represents the magnitude in the $r$-filter at rest. 
We have also calculated $k$~and evolutionary corrections which will be published separately.

\begin{figure*}
 \includegraphics[width=0.49\textwidth]{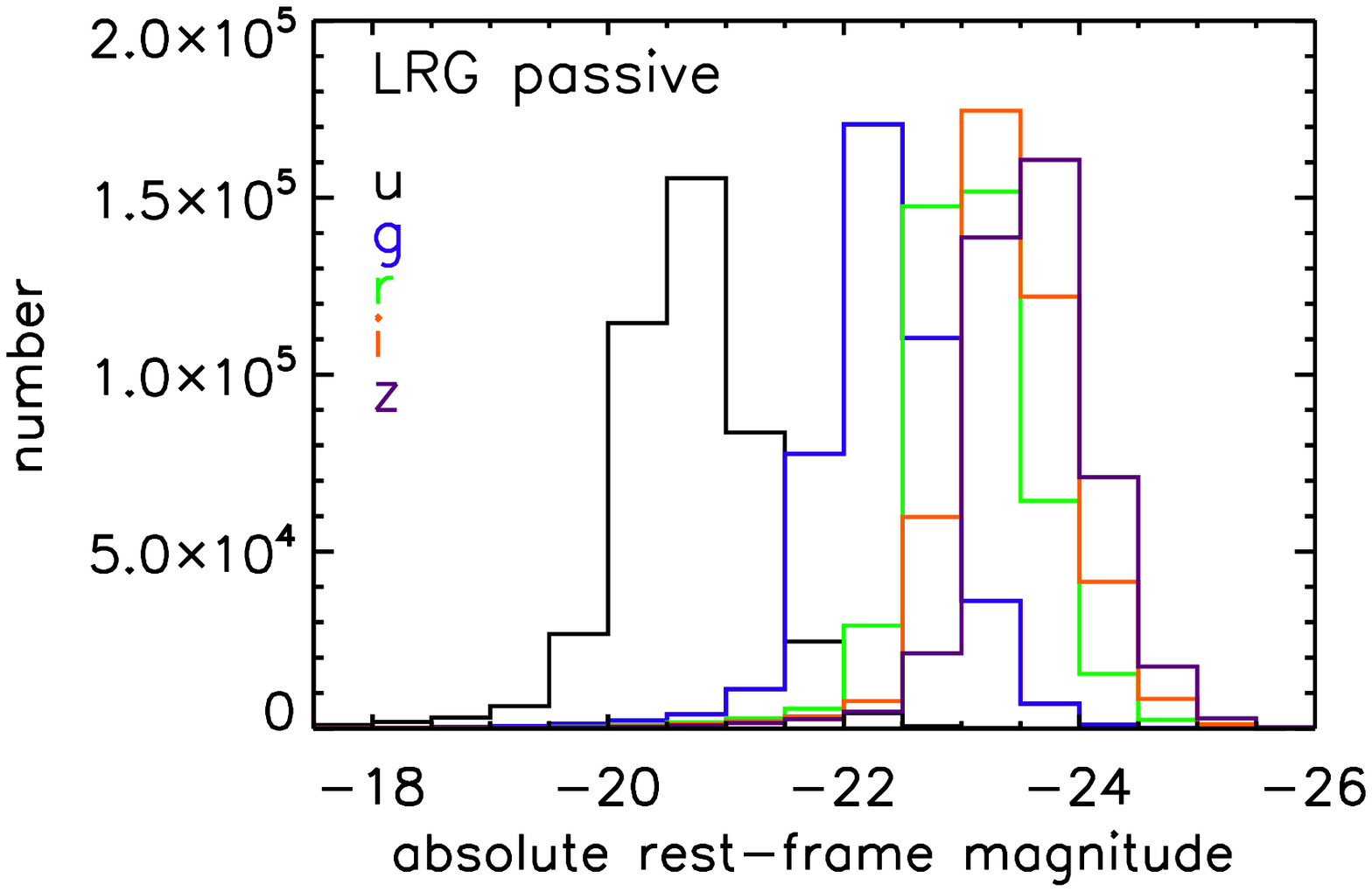}
 \includegraphics[width=0.49\textwidth]{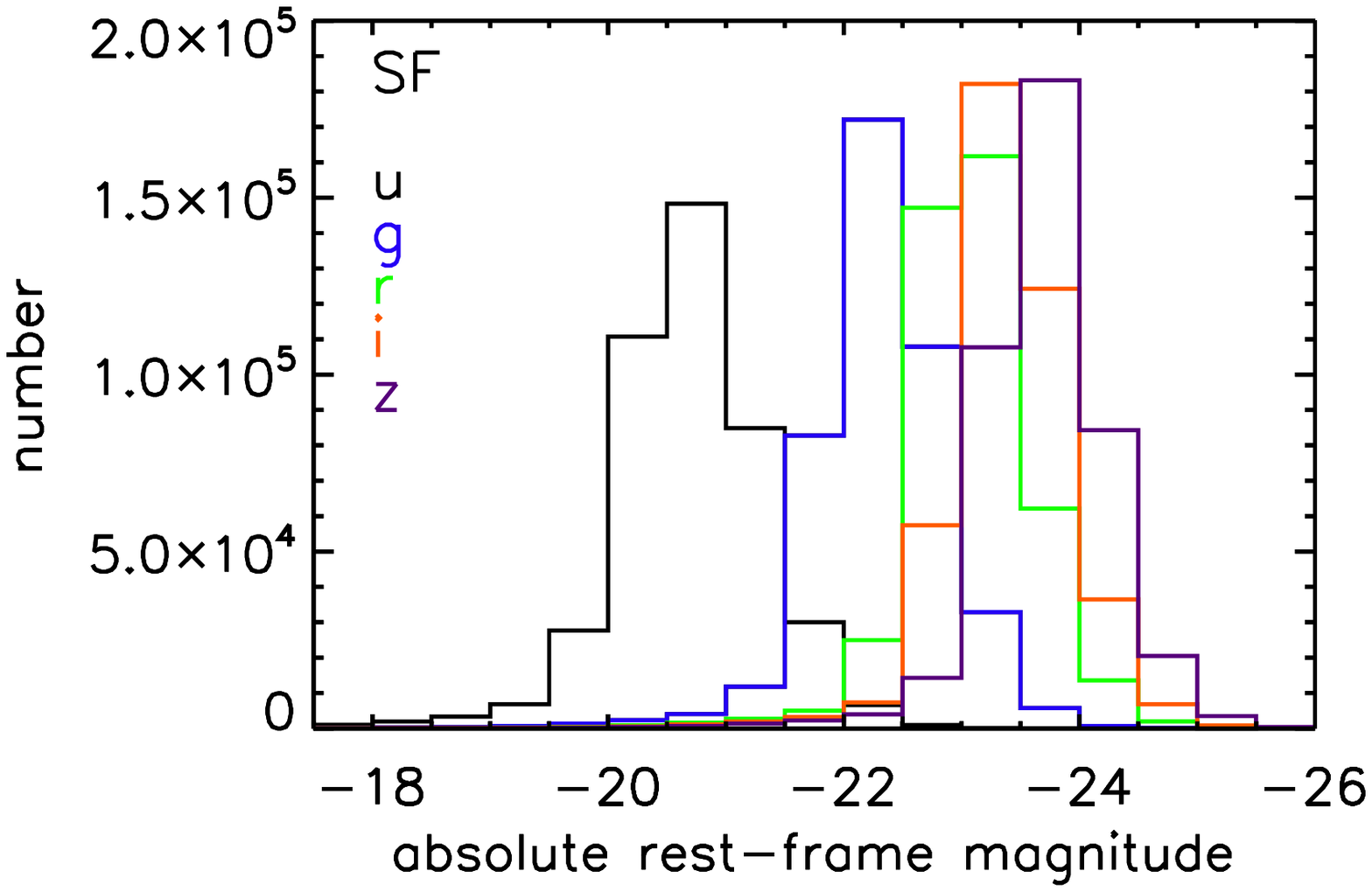}
  \caption{Modelled rest-frame magnitudes of BOSS (CMASS and LOWZ) galaxies in $u,g,r,i,z$~(labelled) for the passive LRG template (left-hand panel), and for the star forming template (right-hand panel).}
  \label{fig:restframemags}
\end{figure*}

The two panels of Figure~\ref{fig:restframemags} show the rest-frame magnitudes of BOSS galaxies according to the passive LRG and the SF template. There is hardly any difference in these results due to the similar age distribution that is obtained independently of the assumed template.
\begin{figure}
 \includegraphics[width=0.49\textwidth]{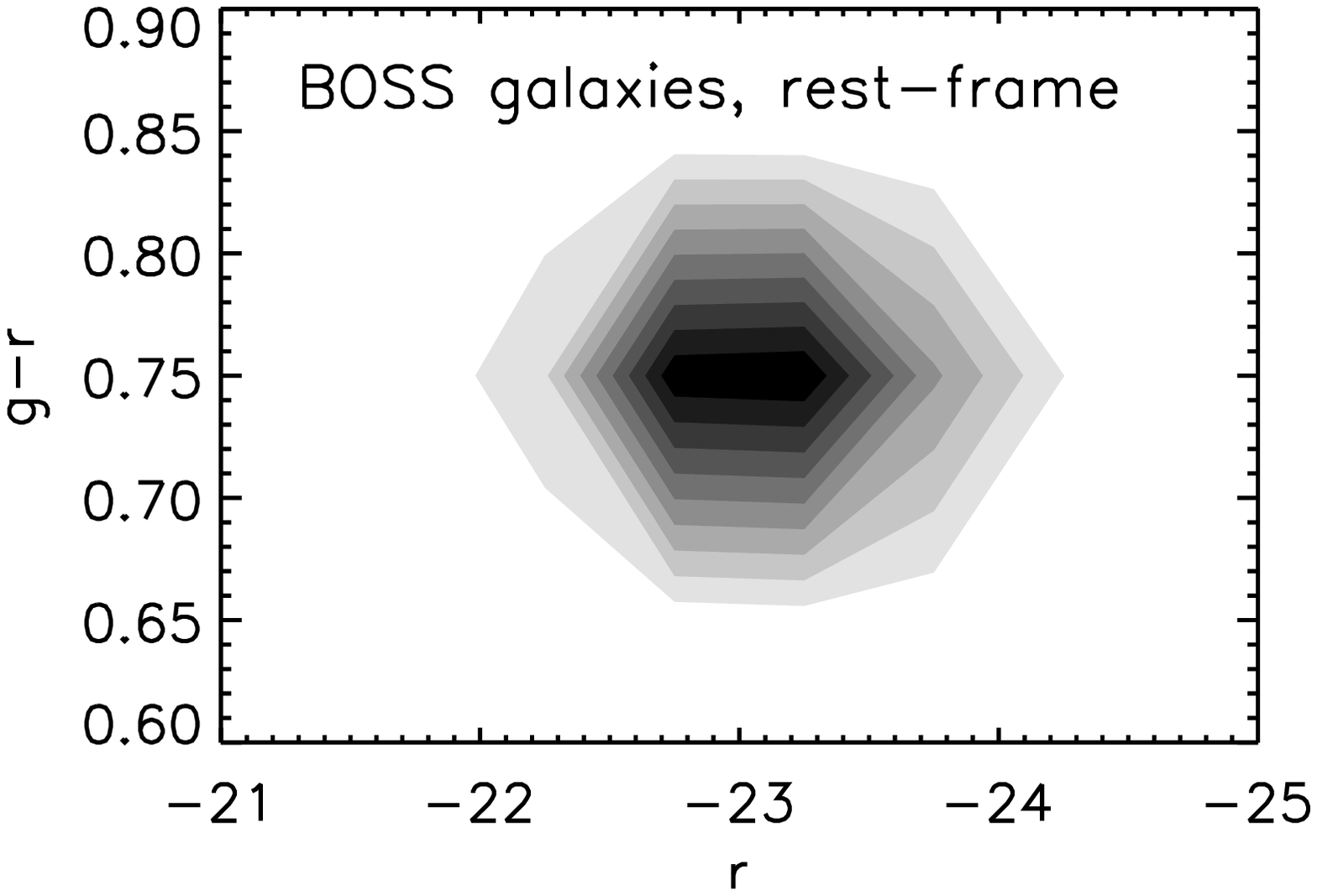}
  \caption{Rest-frame colourr-magnitude diagram of BOSS (CMASS and LOWZ) galaxies for the passive LRG template.}
  \label{fig:cmrest}
\end{figure}

Finally, Figure~\ref{fig:cmrest} shows the rest-frame $g-r$~vs $r$~ colour magnitude diagram for BOSS galaxies. The uniformity of the sample is reflected in a galaxy population spanning a narrow intrinsic colour and magnitude range.

\section{Effect of stellar mass derivation on the stellar mass function.}
\label{magskcorr}

In this Appendix we report the results of experiments in which we varied the stellar mass templates and priors and assessed the effect on the mass function. Each plot should be compared to Figure~\ref{fig:massfunc}. 

\begin{figure*}
  \includegraphics[width=\textwidth]{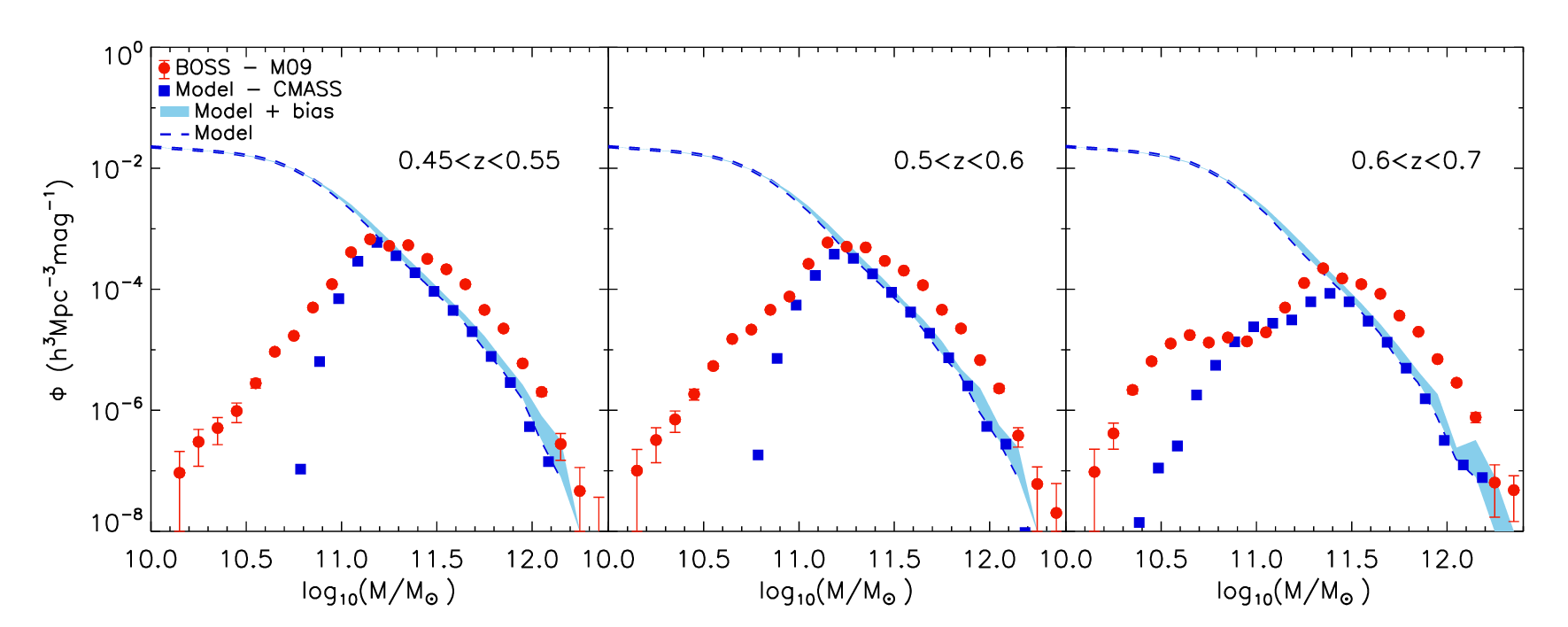}
  \caption{As in Figure~\ref{fig:massfunc}, but without correcting the stellar masses of blue galaxies for the mass offset found for mock galaxies and displayed in Figure~\ref{fig:mockSF}.}
  \label{fig:massfuncnocorr}
\end{figure*}

Figure~\ref{fig:massfuncnocorr} displays the stellar mass function of CMASS galaxies (red points with errors), where we have not applied the correction of $+0.25$ dex to the stellar masses obtained for star forming galaxies (Figure~15, Section 4.1). As can be appreciated from the comparison between the two mass function plots, this creates a discrepancy at 
$M^{*}\sim10^{11}M_{\odot}$, which increases towards higher redshift (due to a higher fraction of galaxies with bluer colours in the selection cut and to generally younger galaxy ages). From the mock experiment, we conclude that this discrepancy is artificial.

\begin{figure*}
  \includegraphics[width=\textwidth]{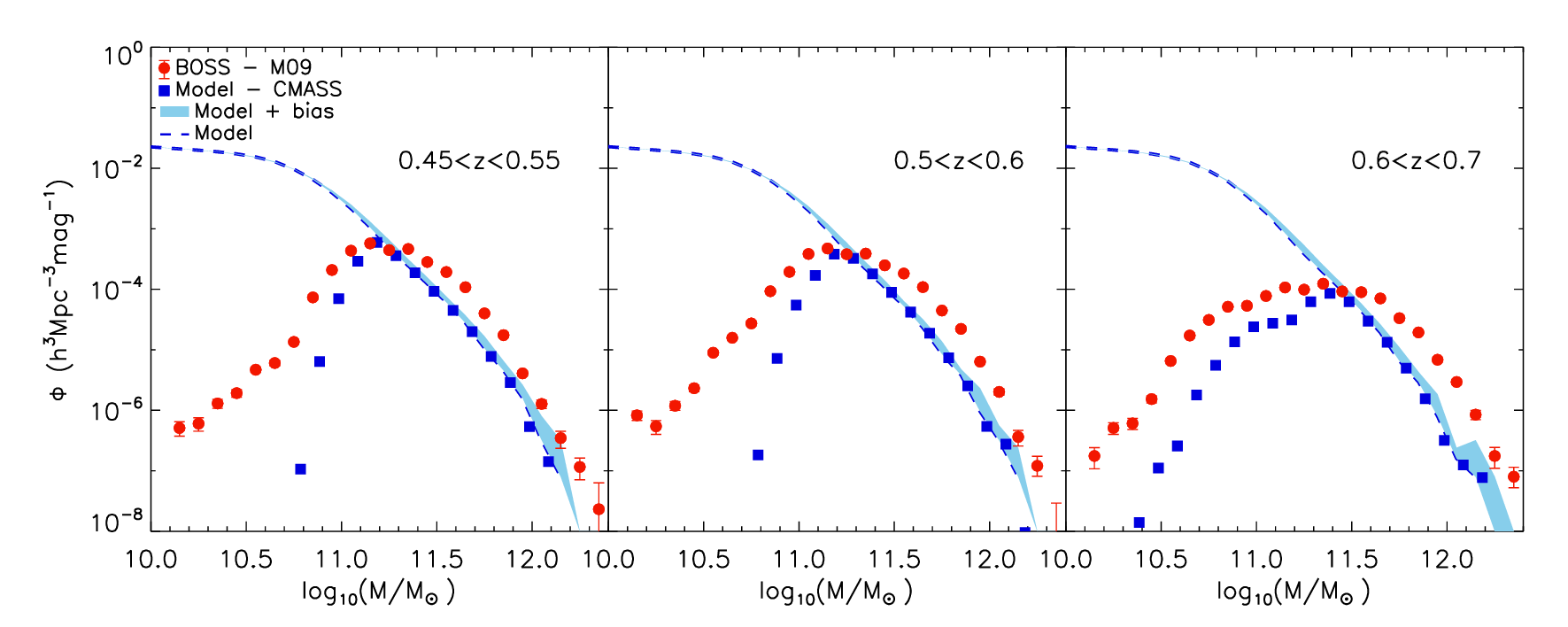}
  \caption{As in Figure~\ref{fig:massfunc}, but without applying a minimum age to the passive LRG template of 3 Gyr (Figure~\ref{fig:mockLRG}).}
  \label{fig:massfuncnotmin}
\end{figure*}

Figure~\ref{fig:massfuncnotmin} shows the mass function where we relax the minimum age constrain of 3 Gyr in the LRG template. In this case, a fraction of galaxies with red $g-i$~colours get fitted ages lower than 3 Gyr, hence a lower stellar mass, which has the effect at slightly shifting the mass distribution towards lower values. This worsens somewhat the comparison between data and models at the lowest mass end. While this option of template fitting is not flawed in principle, in practice it gives underestimated stellar masses for mock galaxies (cf. Figure~\ref{fig:mockLRG}). Based on this, we prefer the option in which a conservative age limit of 3 Gyr is applied.
\begin{figure*}
  \includegraphics[width=\textwidth]{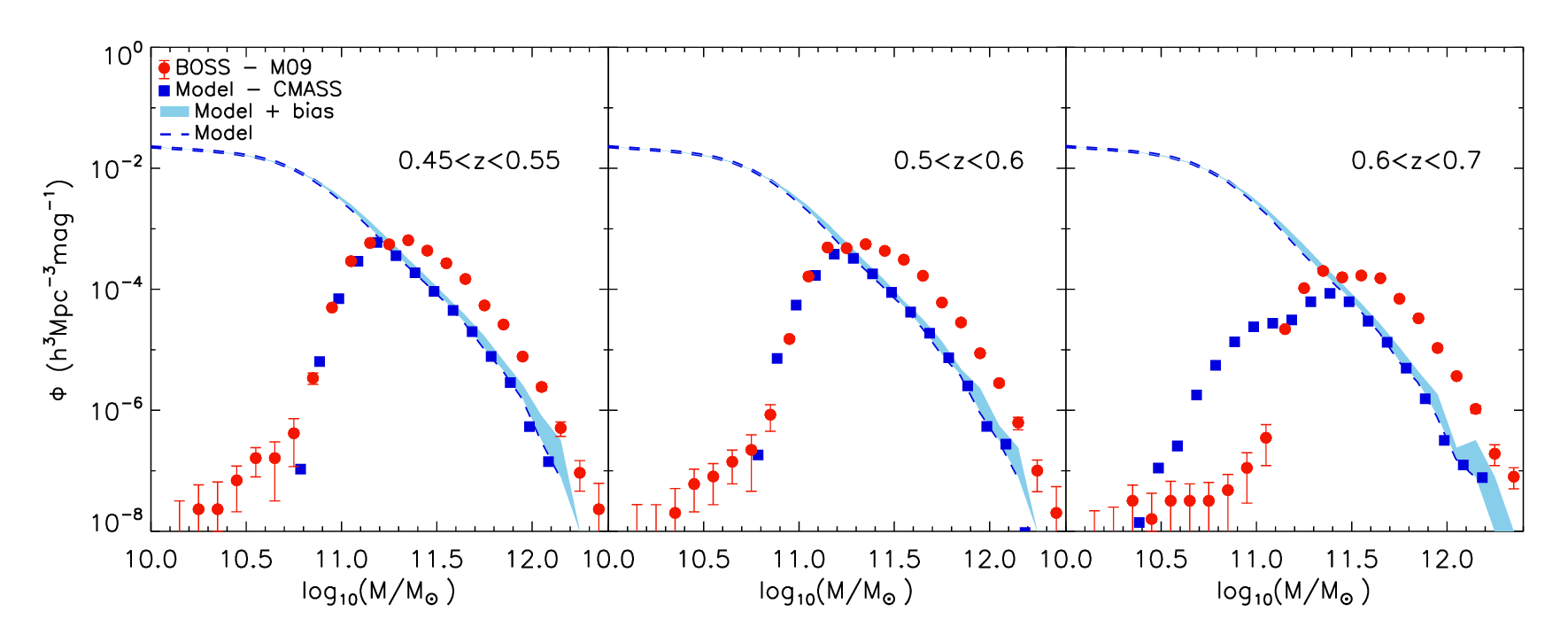}
  \caption{As in Figure~\ref{fig:massfunc}, but using the passive LRG template with minimum age of 3 Gyr for all galaxies independently of their colours.}
  \label{fig:massfunconlylrg}
\end{figure*}

Finally, Figure~\ref{fig:massfunconlylrg} shows the case in which the LRG passive template (with a minimum age constrain)  is used for all galaxies independently of their colours. This gives higher masses to the bluer galaxies which creates a sizable discrepancy at $M^{*}\sim10^{11}~\\M_{\odot}$. The results from Figure~\ref{fig:mockSF} suggest that the use of this template with minimum age of 3 Gyr overestimate the stellar mass of mock galaxies with $g-i\lapprox~2.35$ (black points), hence we conclude that this discrepancy is artificial.

These examples emphasise how crucial the calculation of galaxy stellar masses is for our understanding of galaxy evolution. The assumptions related to the galaxy star formation histories affect the comparison between models and data possibly altering the conclusions.

Note also that - in the case of the BOSS sample which contains a large number of intrinsically red and passive galaxies which are the most massive ones - the assumptions for the bluest or youngest galaxies do not alter the comparison at the high mass end.

\section{Observed-frame colour-magnitude diagrams of BOSS galaxies.}
\label{magskcorr}

Several observed-frame colour magnitude diagrams for BOSS galaxies are displayed in the following figures, which are analogous to  Figure~16. The same conclusions as in Section~5.3 can be drawn from these plots.
\begin{figure*}
  \includegraphics[width=\textwidth]{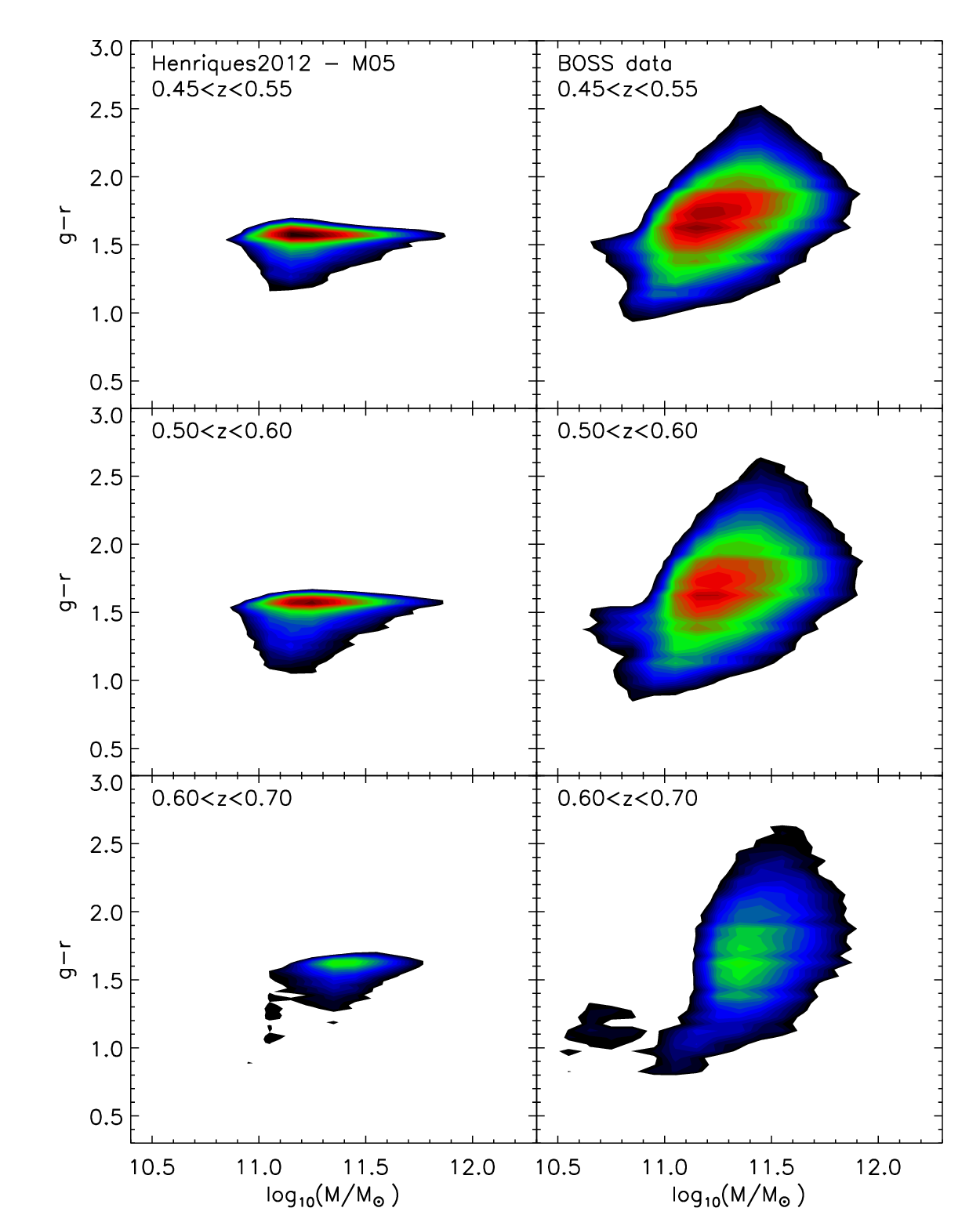}
  \caption{$g-r$~observer-frame colour vs stellar mass for BOSS/CMASS galaxies and semi-analytic models, as in Figure~\ref{fig:colmass}. }
\end{figure*}
\begin{figure*}
 \includegraphics[width=\textwidth]{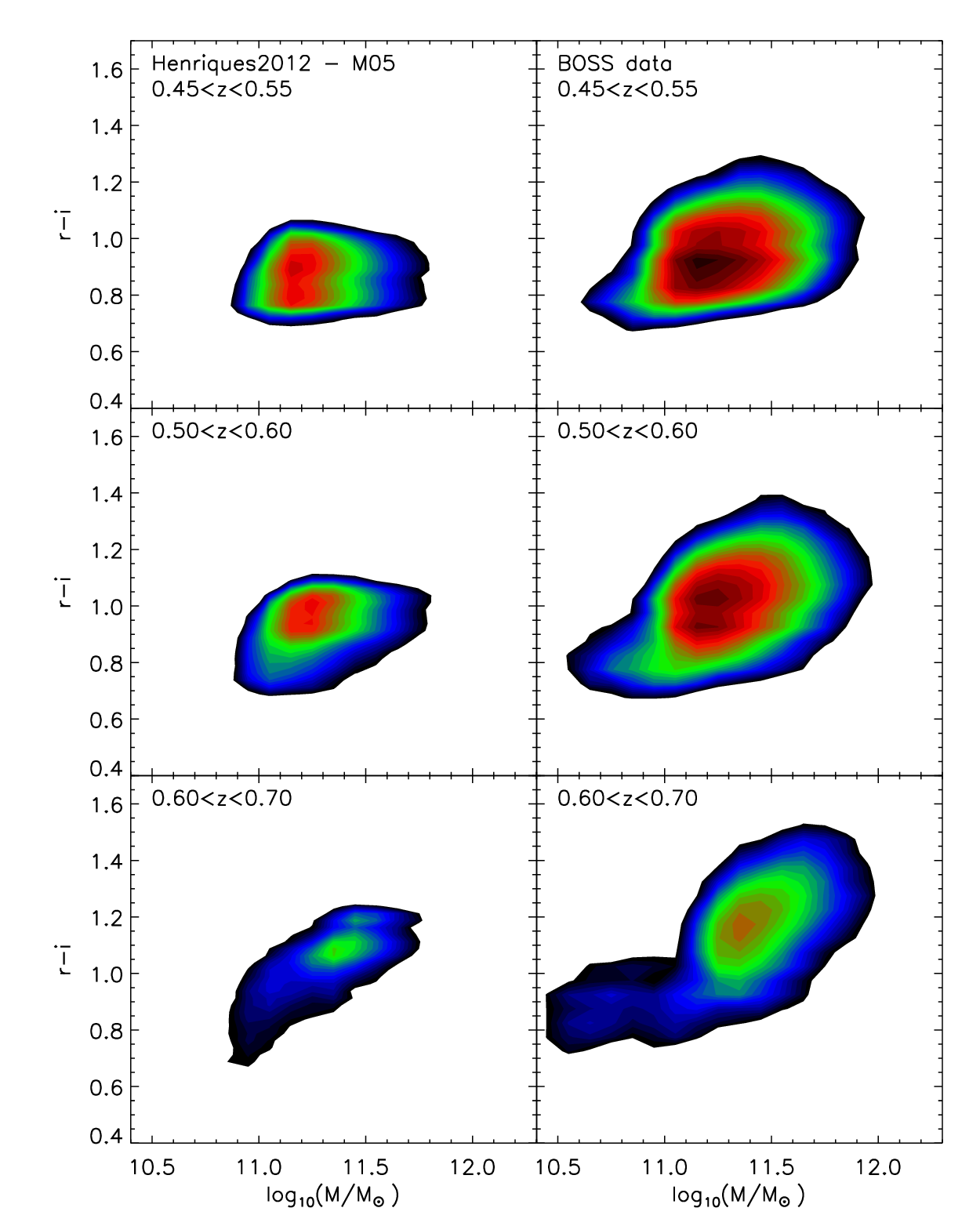}
 \caption{$r-i$~observer-frame colour vs stellar mass for BOSS/CMASS galaxies and semi-analytic models, as in Figure~\ref{fig:colmass}. }
\end{figure*}
\begin{figure*}
  \includegraphics[width=\textwidth]{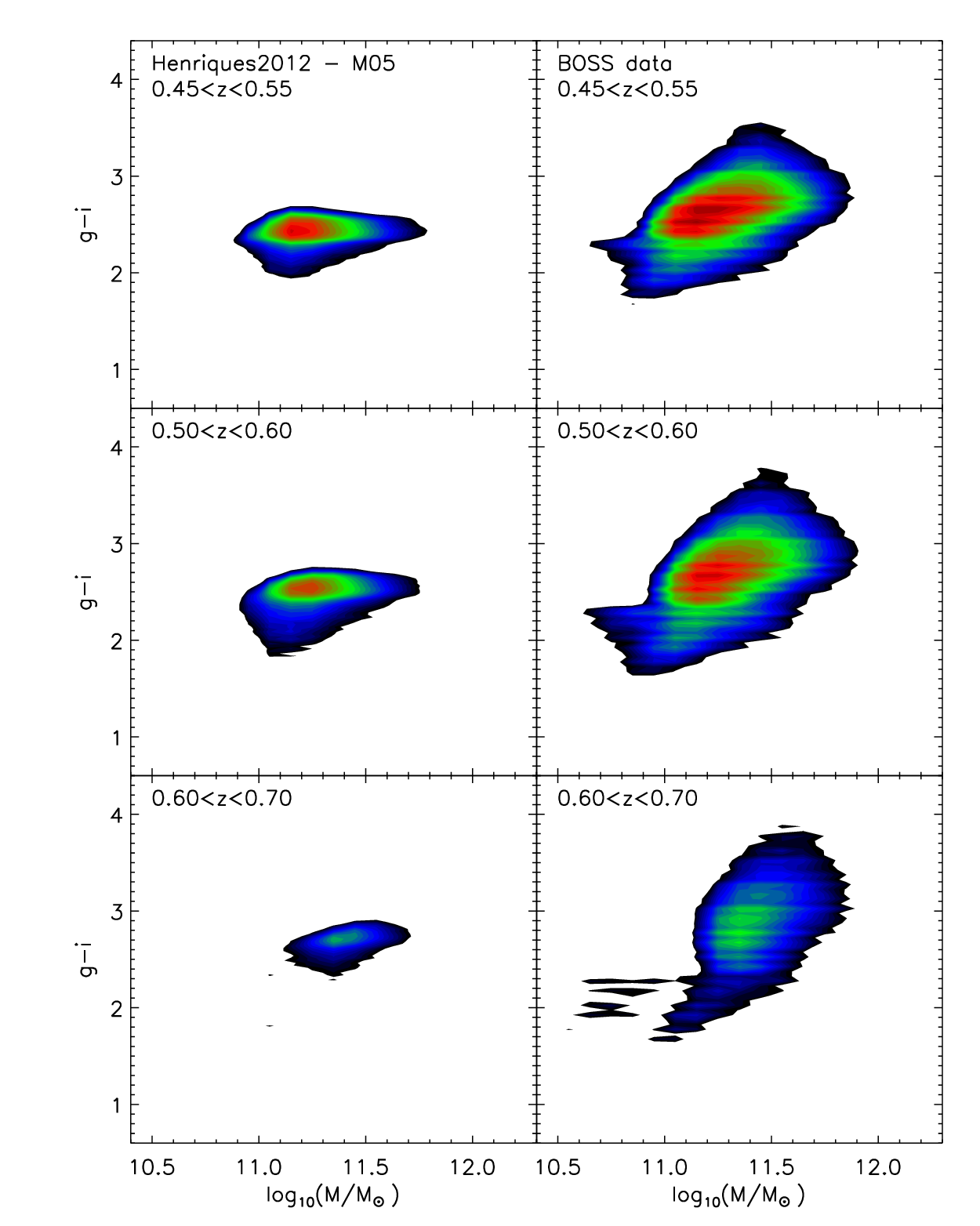}
    \caption{$g-i$~observer-frame colour vs stellar mass for BOSS/CMASS galaxies and semi-analytic models, as in Figure~\ref{fig:colmass}. }
\end{figure*}
\begin{figure*}
  \includegraphics[width=\textwidth]{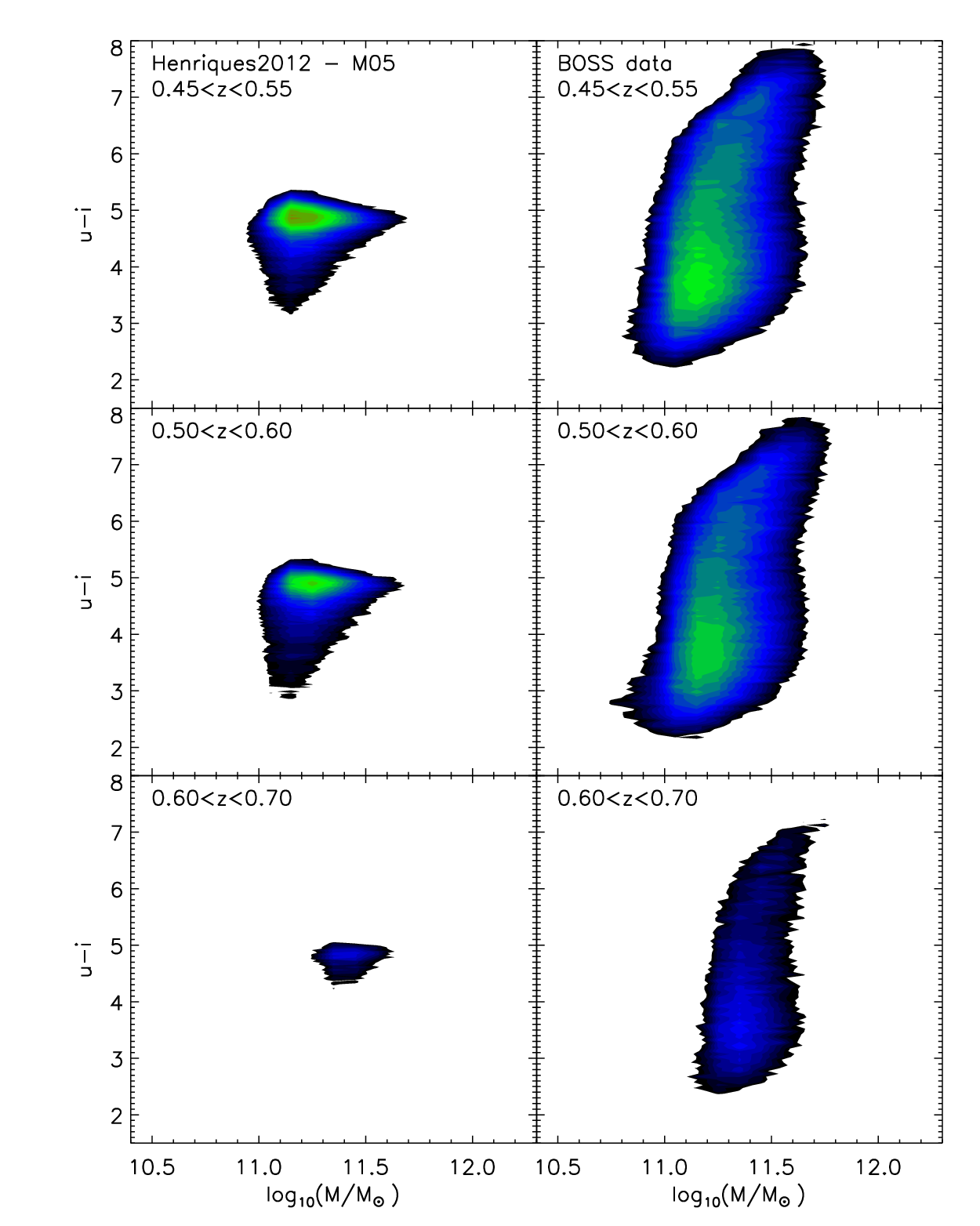}
    \caption{$u-i$~observer-frame colour vs stellar mass for BOSS/CMASS galaxies and semi-analytic models, as in Figure~\ref{fig:colmass}. }
\end{figure*}
\section{Tabulated mass function}
\label{mftable}
\begin{table*}
  \centering
  \begin{minipage}{170mm}
    \caption{Stellar mass function of BOSS galaxies in three redshift bins (as plotted in Figure 19.)}
    \label{tab:mf}
    \begin{tabular}{@{}ccrccc}
      \hline
redshift & $\log M^{*}/M_{\odot}$ &  $N_{\rm gal}$ & $\Phi (h^{3}Mpc^{-3}mag^{-1}$) & $\sigma + $\footnote{Quoted uncertainties include Poisson errors and errors on data-derived $M^{*}$, see Section~5.2.}    & ${\sigma -}^{a}$ \\
\hline
\hline
$0.45 - 0.55$ 
  &     9.95  &   2  & $ 4.627\cdot 10^{-8}$ & $6.150\cdot 10^{-8}$ & $ 3.060\cdot 10^{-8}$\\
  &   10.05  &   7  & $ 1.619\cdot 10^{-7}$ & $ 8.753\cdot 10^{-8}$  & $ 6.010\cdot 10^{-8}$ \\
  &   10.15  &   2  & $ 4.623\cdot 10^{-8}$ & $ 6.149\cdot 10^{-8}$  & $ 3.060\cdot 10^{-8}$  \\
  &   10.25  &   4  & $ 9.254\cdot 10^{-8}$ & $ 7.355\cdot 10^{-8}$  & $ 4.480\cdot 10^{-8}$  \\
  &   10.35  &   6  & $ 1.388\cdot 10^{-7}$ & $ 8.324\cdot 10^{-8}$ &$ 5.547\cdot 10^{-8}$ \\   
  &   10.45  & 10  & $ 2.313\cdot 10^{-7}$  & $ 9.898\cdot 10^{-8}$ & $ 7.224\cdot 10^{-8}$\\
  &   10.55  & 18  & $ 4.164\cdot 10^{-7}$  & $ 9.815\cdot 10^{-8}$  & $ 9.815\cdot 10^{-8}$\\
  &   10.65  & 37  & $ 8.559\cdot 10^{-7}$ & $ 1.407\cdot 10^{-7}$ & $ 1.407\cdot 10^{-7}$\\
  &   10.75  & 93  & $ 2.151\cdot 10^{-6}$ & $ 2.231\cdot 10^{-7}$ & $ 2.231\cdot 10^{-7}$\\
  &   10.85  &  391  & $9.045 \cdot 10^{-6}$ & $4.574\cdot 10^{-7}$  & $4.574\cdot 10^{-7}$ \\
  &   10.95  &  2656 & $6.144\cdot 10^{-5}$  & $1.192\cdot 10^{-6}$ & $1.192\cdot 10^{-6}$ \\
  &   11.05  &  13936 & $3.224\cdot 10^{-4}$ & $2.731\cdot 10^{-6}$  & $2.731\cdot 10^{-6}$ \\
  &   11.15  &  26446 &   $6.120\cdot 10^{-4}$  & $3.762\cdot 10^{-6} $ & $   3.762\cdot 10^{-6}$ \\
  &   11.25  &  23258  &   $5.380\cdot 10^{-4}$  & $  3.528\cdot 10^{-6} $ & $   3.528\cdot 10^{-6}$ \\
  & 11.35     &   25919  &   $5.996\cdot 10^{-4}$ & $  3.724\cdot 10^{-6} $ &  $ 3.724\cdot 10^{-6}$ \\
  & 11.45     &  16346  &   $3.781\cdot 10^{-4}$ & $  2.958\cdot 10^{-6} $&  $  2.958\cdot 10^{-6}$ \\
  & 11.55     &  12119 &    $2.804\cdot 10^{-4}$ & $  2.547\cdot 10^{-6} $ & $  2.547\cdot 10^{-6}$ \\
  & 11.65      &   6851 &    $1.585 \cdot 10^{-4}$  &  $  1.915\cdot 10^{-6} $&  $  1.915\cdot 10^{-6}$ \\
  & 11.75     &   2678 &    $6.193 \cdot 10^{-5}$ &  $ 1.197\cdot 10^{-6}  $&  $ 1.197\cdot 10^{-6}$ \\
  & 11.85     &   1361 &   $3.148 \cdot 10^{-5} $  & $  8.534\cdot 10^{-7} $ & $  8.534\cdot 10^{-7}$ \\
  & 11.95     &   383 &   $8.860 \cdot 10^{-6}$  & $  4.527\cdot 10^{-7} $ &  $ 4.527\cdot 10^{-7}$ \\
  & 12.05     &  134  &   $3.100 \cdot 10^{-6}$ &  $ 2.678\cdot 10^{-7} $ &  $ 2.678\cdot 10^{-7}$ \\
  & 12.15     &   22  &   $5.089 \cdot 10^{-7}$ &  $ 1.085\cdot 10^{-7} $ & $  1.085\cdot 10^{-7}$ \\
  & 12.25     &   10  &   $2.313 \cdot 10^{-7}$ &  $ 9.898\cdot 10^{-8} $ & $  9.898\cdot 10^{-8}$ \\
  & 12.35     &   4  &   $9.254 \cdot 10^{-8}$ &  $ 7.355\cdot 10^{-8} $ & $  4.480\cdot 10^{-8}$ \\
   \hline
 $0.5 - 0.6$  
   &   10.05  &   3  & $ 6.034\cdot 10^{-8}$ & $ 5.907\cdot 10^{-8}$  & $ 3.336\cdot 10^{-8}$ \\
   &   10.15  &   1  & $ 2.011\cdot 10^{-8}$ & $ 4.672\cdot 10^{-8}$  & $ 1.742\cdot 10^{-8}$  \\
   &   10.25  &   2  & $ 4.023\cdot 10^{-8}$ & $ 5.347\cdot 10^{-8}$  & $ 2.661\cdot 10^{-8}$  \\
   &   10.35  &   7  & $ 1.408\cdot 10^{-7}$ & $ 7.611\cdot 10^{-8}$ &$ 5.226\cdot 10^{-8}$ \\   
   &   10.45  & 11  & $ 2.213\cdot 10^{-7}$  & $ 6.671\cdot 10^{-8}$ & $ 6.671\cdot 10^{-8}$\\
   &   10.55  & 31  & $ 6.236\cdot 10^{-7}$  & $ 1.120\cdot 10^{-7}$  & $ 1.120\cdot 10^{-8}$\\
   &   10.65  & 61  & $ 1.227\cdot 10^{-6}$ & $ 1.571\cdot 10^{-7}$ & $ 1.571\cdot 10^{-7}$\\
   &  10.75   &   188  & $ 3.782\cdot 10^{-6}$ &  $2.758 \cdot 10^{-7}$ &  $2.758 \cdot 10^{-7}$\\
   &  10.85   &   575 & $ 1.157 \cdot 10^{-5}$ &  $ 4.823 \cdot 10^{-7}$ &  $4.823 \cdot 10^{-7}$\\
   &   10.95   & 1607 &  $3.232\cdot 10^{-5}$ &  $8.064\cdot 10^{-7}$  & $8.064\cdot 10^{-7}$\\
   &   11.05    &   9912  & $ 1.994 \cdot 10^{-4}$  & $2.003\cdot 10^{-6}$ & $ 2.003\cdot 10^{-6}$\\
  & 11.15  &   26898 &  $5.411\cdot 10^{-4}$ & $3.299\cdot 10^{-6}$ &  $3.299\cdot 10^{-6}$\\
  & 11.25   &  25827  & $ 5.195\cdot 10^{-4}$   & $ 3.233\cdot 10^{-6}$  & $3.233\cdot 10^{-6}$\\
  & 11.55   &  12584  & $2.531\cdot 10^{-4}$  & $2.256\cdot 10^{-6}$  & $2.256\cdot 10^{-6}$\\
  & 11.65   &   7550  & $1.519\cdot 10^{-4}$  & $1.748\cdot 10^{-6}$  & $1.748\cdot 10^{-6}$\\
  & 11.75   &   3207 &  $6.451\cdot 10^{-5}$  & $1.139\cdot 10^{-6}$  & $1.139\cdot 10^{-6}$\\
  & 11.85   &  1603  & $3.224\cdot 10^{-5}$  & $8.054\cdot 10^{-7}$  & $ 8.054\cdot 10^{-7}$\\
  & 11.95   &   482  & $9.695\cdot 10^{-6}$  & $4.416\cdot 10^{-7}$  & $ 4.416\cdot 10^{-7}$\\
  & 12.05   &  166  & $3.339\cdot 10^{-6}$  & $2.592\cdot 10^{-7}$  & $ 2.592\cdot 10^{-7}$\\
  & 12.15   &   36  & $7.241\cdot 10^{-7}$  & $1.207\cdot 10^{-7}$  & $ 1.207\cdot 10^{-7}$\\    
  & 12.25   &   8 &   $1.610 \cdot 10^{-7}$ &  $ 7.962\cdot 10^{-8} $ & $  5.599\cdot 10^{-8}$ \\
  & 12.35   &   4  &   $8.046 \cdot 10^{-8}$ &  $ 6.395\cdot 10^{-8} $ & $  6.395\cdot 10^{-8}$ \\
\hline
  $0.6 - 0.7$  
  &   10.05  &   1  & $ 1.595\cdot 10^{-8}$ & $ 3.705\cdot 10^{-8}$  & $ 1.381\cdot 10^{-8}$ \\
  &   10.15  &   1  & $ 1.595\cdot 10^{-8}$ & $ 3.705\cdot 10^{-8}$  & $ 1.381\cdot 10^{-8}$  \\
  &  10.25  &   3  & $ 4.785\cdot 10^{-8}$ & $ 4.683\cdot 10^{-8}$  & $ 2.645\cdot 10^{-8}$  \\
  &  10.35  &   5  & $ 7.975\cdot 10^{-8}$ & $ 5.419\cdot 10^{-8}$ &$ 3.476\cdot 10^{-8}$ \\   
  &  10.45  & 17  & $ 2.711\cdot 10^{-7}$  & $ 6.576\cdot 10^{-8}$ & $ 6.576\cdot 10^{-8}$\\
  &  10.55    &   73  & $1.164\cdot 10^{-6} $ & $1.363\cdot 10^{-7} $ & $1.363\cdot 10^{-7}$\\
  &  10.65   &  299  & $4.769\cdot 10^{-6} $ & $ 2.758\cdot 10^{-7} $ & $ 2.758\cdot 10^{-7}$\\
  &  10.75   &  627 & $ 1.000\cdot 10^{-5} $ & $ 3.994\cdot 10^{-7} $ & $ 3.994\cdot 10^{-7}$\\
  &  10.85   &  1042  & $1.662e\cdot 10^{-5} $ & $ 5.148\cdot 10^{-7} $ & $ 5.148\cdot 10^{-7}$\\
  &  10.95   &  868  & $1.384\cdot 10^{-5} $ & $ 4.699\cdot 10^{-7}  $ & $ 4.699\cdot 10^{-7}$\\
  &  11.05   &  1043  & $1.663\cdot 10^{-5} $ & $ 5.151\cdot 10^{-7}  $ & $ 5.151\cdot 10^{-7}$\\
  &  11.15   &  2250  & $3.588\cdot 10^{-5} $ & $ 7.565\cdot 10^{-7}  $ & $ 7.565\cdot 10^{-7}$\\
  &  11.25   & 7545 & $ 1.203\cdot 10^{-4} $ & $ 1.385\cdot 10^{-6}  $ & $ 1.385\cdot 10^{-6}$\\
  &  11.35   &  14158 &  $2.258\cdot 10^{-4} $ & $ 1.897\cdot 10^{-6} $ & $ 1.897\cdot 10^{-6}$\\
  &  11.45   & 10136 &  $1.617\cdot 10^{-4} $ & $ 1.606\cdot 10^{-6} $ & $ 1.606\cdot 10^{-6}$\\
  &  11.55   & 8445 &  $1.347\cdot 10^{-4} $ & $ 1.466\cdot 10^{-6} $ & $ 1.466\cdot 10^{-6}$\\
  &  11.65   & 5972 &  $9.525\cdot 10^{-5} $ & $ 1.232\cdot 10^{-6} $ & $ 1.232\cdot 10^{-6}$\\
  &  11.75   & 2881  & $4.595\cdot 10^{-5} $ & $ 8.561\cdot 10^{-7} $ & $ 8.561\cdot 10^{-7}$\\
  &  11.85   & 1688  & $2.692\cdot 10^{-5} $ & $ 6.553\cdot 10^{-7} $ & $ 6.553\cdot 10^{-7}$\\
  &  11.95   &  609  & $9.713\cdot 10^{-6} $ & $ 3.936\cdot 10^{-7} $ & $ 3.936\cdot 10^{-7}$\\
  &  12.05   & 222  & $3.541\cdot 10^{-6} $ & $ 2.376\cdot 10^{-7} $ & $ 2.376\cdot 10^{-7}$\\
  & 12.15   &  61  & $9.729\cdot 10^{-7}  $ & $ 1.246\cdot 10^{-7} $ & $ 1.246\cdot 10^{-7}$\\
  & 12.25   &  17 &   $2.711 \cdot 10^{-7}$ &  $ 6.576\cdot 10^{-8} $ & $  6.576\cdot 10^{-8}$ \\
  & 12.35   &   8  &   $1.276 \cdot 10^{-7}$ &  $ 6.313\cdot 10^{-8} $ & $  4.440\cdot 10^{-8}$ \\
  & 12.45   &   1  &   $1.595 \cdot 10^{-8}$ &  $ 3.705\cdot 10^{-8} $ & $  1.381\cdot 10^{-8}$ \\
  & 12.55   &   1  &   $1.595 \cdot 10^{-8}$ &  $ 3.705\cdot 10^{-8} $ & $  1.381\cdot 10^{-8}$ \\
      \hline
          \end{tabular}
  \end{minipage}
\end{table*}

\bsp

\label{lastpage}

\end{document}